%% file: arxiv.tex
\documentclass[letterpaper,10pt]{article}
\usepackage[T1]{fontenc}
\usepackage{parskip}
\usepackage[font={small}]{caption}
\usepackage[hidelinks]{hyperref}
\usepackage[margin=2.5cm]{geometry}
\usepackage{times}
\usepackage{amsmath,amssymb,amsthm}
\usepackage[affil-it]{authblk}
\usepackage{listings}
\usepackage[round]{natbib}

\usepackage[linesnumbered]{algorithm2e}
\usepackage{graphicx}
\usepackage{subcaption}

\usepackage{tikz}
\usepackage{pgfplots}
\usepackage{grffile}
\pgfplotsset{compat=1.13}
\usetikzlibrary{plotmarks}
\usetikzlibrary{arrows.meta}
\usepgfplotslibrary{patchplots}

\newlength\figureheight
\newlength\figurewidth

\date{}

\DontPrintSemicolon

\renewcommand{\mid}{\,\vert\,}
\newcommand{\farg}{\,\cdot\,}
\newcommand{\ii}{{n}}
\newcommand{\ji}{{j}}
\newcommand{\ki}{{k}}
\newcommand{\anc}{a}
\newcommand{\Prb}[1]{\mathbb{P}\left({#1}\right)}

\newcommand{\psx}[1]{\{#1\}_{^{n=1}}^{_{N_x}}}
\newcommand{\psth}[1]{\{#1\}_{^{j=1}}^{_{N_\theta}}}
\newcommand{\pfv}{\xi}

\newcommand{\SMCt}{SMC\textsuperscript{2}\xspace}

\title{Learning of state-space models with highly informative observations: a tempered Sequential Monte Carlo solution}
\author{Andreas Svensson\thanks{\url{andreas.svensson@it.uu.se}}}
\author{Thomas B. Sch\"{o}n}
\author{Fredrik Lindsten}
\affil{Department of Information Technology, Uppsala University}

\begin{document}

\maketitle

\textbf{Please cite this version:}

Andreas Svensson, Thomas B. Sch\"on, Fredrik Lindsten, 2018. Learning of state-space models with highly informative observations: a tempered Sequential Monte Carlo solution. In \textit{Mechanical Systems and Signal Processing}, Volume 104, pp. 915-928.

\begin{center}
\begin{minipage}{.9\linewidth}
\begin{lstlisting}[breaklines,basicstyle=\small\ttfamily]
@article{SvenssonSL:2018,
author    = {Svensson, Andreas and Sch\"on, Thomas B. and Lindsten, Fredrik},
title     = {Learning of state-space models with highly informative observations: a tempered Sequential {Monte} {Carlo} solution},
journal   = {Mechanical Systems and Signal Processing},
volume    = {104},
year      = {2018},
pages     = {915--928},
}
\end{lstlisting}
\end{minipage}
\end{center}

\vspace{5em}

\begin{abstract}
Probabilistic (or Bayesian) modeling and learning offers interesting possibilities for systematic representation of uncertainty using probability theory. However, probabilistic learning often leads to computationally challenging problems. Some problems of this type that were previously intractable can now be solved on standard personal computers thanks to recent advances in Monte Carlo methods. In particular, for learning of unknown parameters in nonlinear state-space models, methods based on the particle filter (a Monte Carlo method) have proven very useful. A notoriously challenging problem, however, still occurs when the observations in the state-space model are highly informative, i.e. when there is very little or no measurement noise present, relative to the amount of process noise. The particle filter will then struggle in estimating one of the basic components for probabilistic learning, namely the likelihood $p($data$|$parameters$)$. To this end we suggest an algorithm which initially assumes that there is substantial amount of artificial measurement noise present. The variance of this noise is sequentially decreased in an adaptive fashion such that we, in the end, recover the original problem or possibly a very close approximation of it. The main component in our algorithm is a sequential Monte Carlo (SMC) sampler, which gives our proposed method a clear resemblance to the \SMCt method. Another natural link is also made to the ideas underlying the approximate Bayesian computation (ABC).
We illustrate it with numerical examples, and in particular show promising results for a challenging Wiener-Hammerstein benchmark problem.

\end{abstract}

\clearpage


\section{Introduction}

Probabilistic (or Bayesian) modeling and learning offers interesting and promising possibilities for a coherent and systematic description of model and parameter \emph{uncertainty} based on probability theory \cite{Peterka:1981,Robert:2001}. The computational tools for probabilistic learning in state-space models have lately been developed. In this paper, we study probabilistic learning based on measured data $\{y_1, \dots, y_T\} \triangleq y_{1:T}$, which we assume to be well described by a nonlinear state-space
model with (almost) no measurement noise,
\begin{subequations}\label{eq:ssm} 
	\begin{align}
	x_{t}\mid (x_{1:t-1},\theta) & \sim f(x_{t}\mid x_{t-1},u_{t-1},\theta),\label{eq:ssm:f}\\
	y_{t} & =g(x_{t}),\label{eq:ssm:g}
	\end{align}
\end{subequations}
with some unknown parameters $\theta\in\Theta$ which we want to learn. The lack
of measurement noise in (\ref{eq:ssm:g}) gives a deterministic mapping $g:\mathsf{X}\mapsto \mathsf{Y}$ from the unobserved states $x_t\in\mathsf{X}$ to the measurement $y_t\in\mathsf{Y}$, on the contrary to (\ref{eq:ssm:f}) which encodes uncertainty about $x_{t}$, mathematically represented as a probability density $f$ over $x_{t}$ conditional on $x_{t-1}$ and possibly an exogenous input $u_{t-1}$. We refer to this uncertainty as \emph{process noise}, but its origin does not have to be a physical noise, but possibly originating from lack of information or model errors. The reasoning and contributions of this paper will be applicable also to the case where the relationship (\ref{eq:ssm:g}) does contain uncertainty, \emph{measurement noise}, but its variance is much smaller than the process noise. As a general term, we refer to the model as having \emph{highly informative observations}. Furthermore, $g$ could 
also be allowed to depend on $\theta$ and $u_{t}$, but we omit that possibility for notational clarity.

Models on the form~\eqref{eq:ssm} may arise in several practical situations, for instance in a mechanical system where the measurements can be made with good precision but some unobserved forces are acting on the system. The situation may also appear if the measurements, yet again, can be made with good precision, but the user's understanding of the physical system is limited, which in the probabilistic framework can be modeled as a stochastic element in $f$.

The model~\eqref{eq:ssm} defines, together with priors on $\theta$, a joint probabilistic model $p(y_{1:T},x_{1:T},\theta)$. Probabilistic learning of the parameters $\theta$ amounts to computing the parameter posterior $p(\theta\mid y_{1:T})$, where we have conditioned on data $y_{1:T}$ and marginalized over all possible states $x_{1:T}$ (we omit the known $u_{1:T}$ to ease the notation). Although conceptually clear, the computations needed are typically challenging, and almost no cases exist that admit closed-form expressions for $p(\theta\mid y_{1:T})$.

For probabilistic learning, Monte Carlo methods have proven useful, as outlined in the accompanying paper \cite{SSM+:2017}. The idea underlying these Monte Carlo methods is to represent the distributions of interest, such as the posterior $p(\theta\mid y_{1:T})$, with samples. The samples can later be used to estimate functions of the parameters $\theta$, such as their mean, variance, etc., as well as making predictions of future outputs $y_{T+1}$, etc. For state-space models, the particle filter is a tailored algorithm for handling the unknown states $x_t$, and in particular to compute an unbiased estimate $z$ of the likelihood 
\begin{equation}
p(y_{1:T}\mid\theta) = \int p(y_{1:T},x_{1:T}\mid\theta)dx_{1:T},\label{eq:likelihood}
\end{equation}
which is a central object in probabilistic learning, see the accompanying paper \cite{SSM+:2017} for a more thorough introduction (or, e.g., \cite{SLD+:2015,KDS+:2015}). The peculiarity in the problem studied in this paper is the (relative) absence of measurement noise in~\eqref{eq:ssm} compared to the process noise level. This seemingly innocent detail is, as we will detail in Section~\ref{sec:challenges}, a show-stopper for the standard algorithms based on the particle filter, since the quality of the likelihood estimate $z$ tends to be very poor if the model has highly informative observations.

The problem with highly informative observations has a connection to the literature on approximate Bayesian computations (ABC, \cite{BZB:2002}), where some observations $y$ are available, as well as a model (not necessarily a state-space model) with some unknown parameters $\theta$. In ABC problems, however, the model is only capable of \emph{simulating} new synthetic observations $\widehat{y}(\theta)$ and the likelihood $p(y\mid\theta)$ cannot be evaluated. The ABC idea is to construct a distance metric between the real observations $y$ and the simulated synthetic observations $\widehat{y}(\theta)$, and take this distance (which becomes a function of $y$ and $\theta$) as a substitute for $p(y\mid\theta)$.
The accuracy of the approximation is controlled by the metric with higher accuracy corresponding to more informative observations, providing a clear link to the present work.

We propose in this paper a novel algorithm for the purpose of learning $\theta$ in~\eqref{eq:ssm}. Our idea is to start the algorithm by assuming that there \emph{is} a substantial amount of measurement noise which mitigates the computational problems, and then gradually decrease this artificial measurement noise variance simultaneously as the parameters $\theta$ are learned. The assumption of artificial measurement noise resembles the ABC methodology. The sequence of gradually decreasing measurement noise variance can be seen as tempering, which we will combine with a sequential Monte Carlo (SMC) sampler \cite{DDJ:2006} to obtain a theoretically sound algorithm  which generates samples from the posterior $p(\theta\mid y_{1:T})$.

In a sense, our proposed algorithm is a combination of the work by \cite{DeanSJP:2014} on ABC for state-space models and the use of SMC samplers for ABC by \cite{DDJ:2012}, resulting in a \SMCt-like algorithm \cite{CJP:2013}.

\section{Background on particle filtering and tempering}

In this section we will provide some background on 
particle filters, Markov chain Monte Carlo (MCMC) and related methods. For a more elaborate introduction, please refer to, e.g., \cite{SSM+:2017,DS:2015,RC:2004}. We will in particular discuss why models on the form~\eqref{eq:ssm} are problematic for most existing methods, and also introduce the notion of tempering. 

\subsection{Particle filtering, PMCMC and \SMCt}\label{sec:intro:mc}

The bootstrap particle filter was presented in the early 1990's \cite{GSS:1993,DJ:2011} as a solution to the state filtering problem (computing $p(x_t\mid y_{1:t})$) in nonlinear state-space models. The idea is to propagate a set of $N_x$ Monte Carlo samples $\psx{x_t^n}$ along the time dimension $t=1, 2, \dots, T$, and for each $t$ the algorithm follows a 3-stage scheme with resampling (sampling ancestor indices $a_t^n$ based on weights $w_{t-1}^n$), propagation (sampling $x_t^n$ from $x_{t-1}^{a_t^n}$ using~\eqref{eq:ssm:f}) and weighting (evaluate the `usefulness' of $x_t^n$ using~\eqref{eq:ssm:g} and store it as the weight $w_t^n$). This algorithm will be given as Algorithm~\ref{alg:pf}, and a more elaborate introduction can be found in \cite{SSM+:2017}. The samples are often referred to as particles, and provide an empirical approximation $\widehat{p}(x_t\mid y_{1:t}) = \frac{1}{N_x} \sum_{n=1}^{N_x} \delta_{x_t^n}(x_t)$ (with $\delta$ the Dirac measure) of the filtering distribution $p(x_t\mid y_{1:t})$. Since the particle filter itself builds on Monte Carlo ideas, the outcome of the algorithm will be different every time the algorithm is run.

\emph{The particle filter itself is only applicable when the state-space model does not contain any unknown parameters}. It has, however, been realized that the particle filter does not only solve the filtering problem, but it can also be used to estimate the likelihood $p(y_{1:T}\mid \theta)$ of a state-space model by using the empirical approximation $\widehat{p}(x_t\mid y_{1:t})$ in~\eqref{eq:likelihood} and hence approximate the integral with a sum. We will denote the obtained estimate with $z$, and it can be shown \cite{DelMoral:2004} that $z$ is in fact an unbiased estimator of the likelihood, $\mathbb{E}[z-p(y_{1:T}\mid\theta)] = \mathbb{E}[z]-p(y_{1:T}\mid\theta) = 0$. That is, unbiased means that the average of the estimate $z$ (if the particle filter algorithm is run many times for the same model, the same parameters and the same data) will be close to the true (but intractable) $p(y_{1:T}\mid \theta)$.

The use of the particle filter as an estimator of the likelihood has opened up possibilities for combining it with another branch of Monte Carlo methods, namely Markov chain Monte Carlo (MCMC). This combination allows for inferring not only unobserved states $x_t$ but \emph{also} unknown parameters $\theta$ in nonlinear state-space models. One such successful idea is to construct a high-level MCMC procedure concerned with $\theta$, and then run the particle filter to estimate the likelihood for different $\theta$. The high-level procedure can be a Metropolis-Hastings algorithm \cite{MRR+:1953}, \cite[Section 5]{SSM+:2017}, essentially an informed random walk in $\Theta$. The Metropolis-Hastings algorithm is constructed such that after sufficiently long time, the trace of the `walk' (the `chain') in $\Theta$ will be samples from the distribution we are interested in, $p(\theta\mid y_{1:T})$.

The original Metropolis-Hastings algorithm assumes that the target distribution can be evaluated exactly. In our state-space learning problem it would mean that the stochastic estimate $z$ from the particle filter would not be sufficient for a valid Metropolis-Hastings algorithm. However, it has lately been shown \cite{AR:2009} that valid algorithms can be constructed based also on stochastic estimates with certain properties, which provides the ground for the particle (marginal) Metropolis-Hastings (PMH, or PMMH, \cite{ADH:2010}) algorithm. We will not go into further details here, but refer to~\cite{SSM+:2017}.

An alternative partner for the particle filter, instead of MCMC, is SMC. Interestingly enough, SMC is a family of methods that has been developed as a generalization of the particle filter. One SMC method is the SMC sampler \cite{DDJ:2006}, which can be employed to handle the unknown $\theta$ instead of Metropolis-Hastings. The SMC sampler will then query the particle filter for likelihood estimates $z$ for different values of $\theta$. The SMC sampler itself is similar to a particle filter, propagating its $N_\theta$ samples $\psth{\theta^\ji}$ through a sequence of distributions ending up in the posterior $p(\theta\mid y)$. The sequence through which the samples of $\theta$ are propagated can be a so-called tempering sequence. With a certain choice of tempering sequence, the nested construction of the particle filter and SMC sampler has been termed \SMCt \cite{CJP:2013}. The method that we propose in this paper bears close resemblance to \SMCt, but makes use of a different tempering sequence.

\subsection{Challenges with highly informative observations}\label{sec:challenges}

The bootstrap particle filter is often used to provide estimates $z$ of the likelihood $p(y_{1:T}\mid\theta)$ in probabilistic learning methods. However, when
there is (almost) no measurement noise relative to the amount of process noise, and thus highly informative observations, these estimates become poor 
due to the importance sampling mechanism inherent in the particle filter. In
the bootstrap particle filter, $N_x$ particles $\psx{x_{t}^{n}}$ are drawn from
(\ref{eq:ssm:f}), and then weighted by evaluating (\ref{eq:ssm:g}). As long as there is at least one particle $x_{t}^{n}$
wich gives a reasonably high probability for the measurement $y_{t}$
(and consequently gets assigned a large weight), the particle filter
will provide a reasonable result, and the more such high-weight particles,
the better (in terms of variance of the estimate $z$).
However, if no samples $x_t$ are drawn under which $y_{t}$ could have been observed with reasonably high probability, the estimate $z$
will be very poor. If there is very little measurement noise but non-negligible process noise in the model, the chance of drawing
any useful particles $x_{t}^{n}$ by simulating the system dynamics is typically small. The problem may
become even more articulated if the bootstrap particle filter is run with a parameter
$\theta$ which does not explain the measurements $y_{1:T}$ well.
The bottom line is that a model with highly informative observations
causes the bootstrap particle filter to provide estimates $z$
with high variance. This is in particular true for values of $\theta$ that
do not explain the measurements well. Considering that high variance of $z$ implies bad performance in the high-level MCMC or SMC sampler for $\theta$, the model~\eqref{eq:ssm} is problematic to learn.

To this end, research has been done on how to improve the particle filter by drawing particles $\psx{x_{t}^{n}}$ not from (\ref{eq:ssm:f}) but instead from a tailored proposal which also depend on $y_{t}$, in order to better adapt to the measurement $y_t$ and make more `well-informed' particle draws. In the interest of a maintained  consistency, the weight update is modified accordingly. Such adaptation is not always simple, but proposed methods include the fully adapted auxiliary particle filter \cite{PS:1999} (only possible for a limited set of model structures), the alive particle filter \cite{DJL+:2015} and the bridging particle filter \cite{DM:2015} (both computationally more costly). In this work, we will not focus on this aspect, but rather on how inference about $\theta$ can be constructed in order to (as far as possible) avoid running the particle filter for models with highly informative observations. Ultimately, our suggested approach could be combined with methods like the fully adapted, alive or bridging particle filter to push the limits even further.

\subsection{Tempering}\label{sec:tempering}

To construct inference algorithms, the computational trick of tempering
(or annealing) has proven useful. The name tempering
was originally used for a certain heat treatment method within metallurgy, but the term
is also used in a figurative sense for a set of computational methods.
The idea is to construct a `smooth' sequence $\{\pi_{p}(\theta)\}_{p=0}^{P}$
starting in a user-chosen initial function $\pi_{0}(\theta)$ and
ending in the target function $\pi_{P}(\theta)$. In our case, these functions are probability densities, and our target is $\pi_{P}(\theta) = p(\theta\mid y_{1:T})$, as illustrated in Figure~\ref{fig:temp1}. 
There are several ways in which such a sequence can be constructed. We do
not have a formal definition of `smooth', but understand it as a sequence
where every adjacent pair $\{\pi_{p}(\theta), \pi_{p+1}(\theta)\}$
are similar in, e.g., total variation sense. By tracking the
evolution from the typically simple and unimodal $\pi_{0}(\theta)$ to the potentially
intricate and multimodal target $\pi_{P}(\theta)$, the risk of getting
stuck in zero-gradient regions or in local optima is reduced, compared to standard
methods starting directly in $\pi_{P}(\theta)=p(\theta\mid y_{1:T})$.

For state-space models, there are several generic choices for constructing tempering sequences ending up in a posterior $p(\theta\mid y_{1:T})$. One choice (with $P=T$) is the data-tempered sequence $\pi_{p}(\theta)=p(\theta\mid y_{1:p})$, which gives a sequence starting in the prior $p(\theta)$ and, by sequentially including one additional measurement $y_{p}$, eventually ending up in the posterior $p(\theta\mid y_{1:T})$. Typically, the landscape of $p(\theta\mid y_{1:p})$ does not change dramatically when including one extra measurement, and smoothness of the sequence is thus ensured. Another choice is found by first noting that $p(\theta\mid y_{1:T})\propto p(y_{1:T}\mid\theta)p(\theta)$, and then making the choice $\pi_{p}(\theta) \propto p(y_{1:T}\mid\theta)^{p/P}p(\theta)$. Such a sequence also starts, with $p=0$, in the prior $p(\theta)$ and ends, with $p=P$, in the posterior $p(\theta\mid y_{1:T})$. We will in this paper, Section~\ref{sec:strategy:temp}, introduce a new tempering sequence that is tailored for state-space models with highly informative observations, inspired by the ABC approach.

\subsection{Using a tempering sequence in an SMC sampler}

\begin{figure}[t]
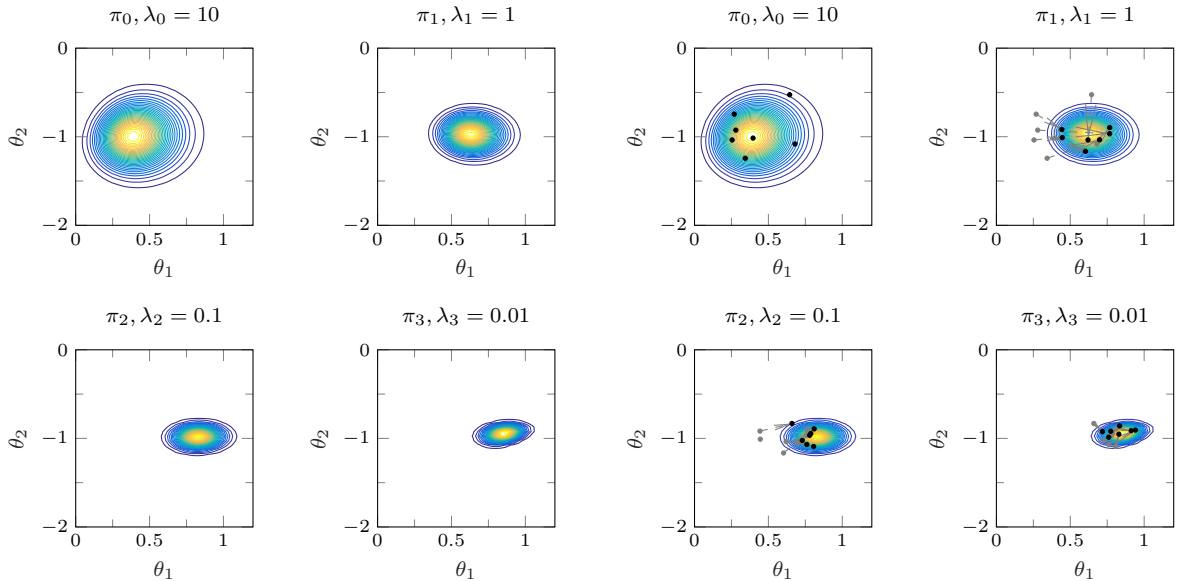

	\centering
	\begin{subfigure}[t]{.48\linewidth}
		\setlength{\figureheight}{.8\linewidth}
		\setlength{\figurewidth}{.8\linewidth}
		\pgfplotsset{
			axis on top,
			label style={font=\scriptsize},
			legend style={inner xsep=1pt,inner ysep=0.5pt,nodes={inner sep=1pt,text depth=0.1em},font=\scriptsize},
			minor y tick num=1,
			minor x tick num=1,
			tick label style={font=\scriptsize},
			every axis/.append style={%
				scaled x ticks=false,
				scaled y ticks=false}
		}
		\centering
		\footnotesize
		\input{tempering1.tex}
		\caption{A tempering sequence shown by level curves for $\pi_p(\theta)$. The tempering sequence used in this figure is the sequence~\eqref{eq:tempp} proposed in this paper, for a linear state-space model with two unknown parameters $\theta_1$ and $\theta_2$. With decreasing $\lambda_{p}$, which is the variance of an artificial measurement noise, tempering is obtained, starting in a distribution with
			a broad support, and ending in a more narrow and peaky distribution.}
		\label{fig:temp1}
	\end{subfigure}
	~
	\begin{subfigure}[t]{.48\linewidth}
		\setlength{\figureheight}{.8\linewidth}
		\setlength{\figurewidth}{.8\linewidth}
		\pgfplotsset{
			axis on top,
			label style={font=\scriptsize},
			legend style={inner xsep=1pt,inner ysep=0.5pt,nodes={inner sep=1pt,text depth=0.1em},font=\scriptsize},
			minor y tick num=1,
			minor x tick num=1,
			tick label style={font=\scriptsize},
			every axis/.append style={%
				scaled x ticks=false,
				scaled y ticks=false}
		}
		\centering
		\footnotesize
		\input{tempering2.tex}
		\caption{The problem of localizing the peak of $\pi_{3}(\theta)$ can be solved
			with an SMC sampler which propagates samples (black dots) through the smooth evolution of the sequence from $p=0$
			to $3$. This is instead of starting to search directly in $\pi_{3}(\theta)$, which would be challenging because of the large `flat' areas. The problem of large uninteresting regions becomes particularly articulated in high dimensional problems.}
		\label{fig:temp2}
	\end{subfigure}
	\caption{An illustration of the tempering idea. The model used here will later be properly introduced as an example in Section~\ref{sec:numex:1}.}
	\label{fig:temp}
	
\end{figure}

A tempering sequence
$\{\pi_{p}(\theta)\}_{p=0}^{P}$ can be used in an SMC sampler
to produce samples from $\pi_{P}(\theta) = p(\theta\mid y_{1:T})$. The idea underlying the SMC sampler is to propagate a set
of $N_{\theta}$ samples $\psth{\theta^{j}}$ along the tempering sequence, and---thanks
to the smoothness of the sequence---gain a high computational efficiency
by generating samples primarily in the most relevant
part of $\Theta$, compared to more basic sampling schemes
such as importance sampling. One version of the SMC sampler is a sequential iteration
of importance sampling and resampling on the sequence $\{\pi_{p}(\theta)\}_{^{p=0}}^{_P}$,
proceeding as follows: samples $\psth{\theta^{j}}$ are initially drawn from
$\pi_{0}(\theta)$, and assigned importance weights $W_1^\ji$ from the ratio $\frac{\pi_{1}(\theta^{j})}{\pi_{0}(\theta^{j})}$.
The samples are then resampled and
moved around in the landscape of $\pi_{1}(\theta)$ with one or a few steps with Metropolis-Hastings. They are then weighted according to $\frac{\pi_{2}(\theta^{j})}{\pi_{1}(\theta^{j})}$, and the procedure is repeated. After $P$ such iterations,
samples from $\pi_{P}(\theta)$ are obtained. An illustration can be found in Figure~\ref{fig:temp2}.

A reader familiar with PMH may understand this use of the SMC sampler as a manager of $N_\theta$ parallel PMH chains, which aborts and duplicates the chains in order to optimize the overall performance\footnote{A subtle but important difference to vanilla PMH is that a PMH chain is typically initialized arbitrarily, run until it converges, and thereafter its transient behavior (the burn-in period) is discarded. In the SMC sampler, however, all chains are `warm-started' thanks to the resampling mechanism, and it is therefore \emph{not} relying on asymptotics to avoid burn-in periods.}.

\section{Solution strategy}\label{sec:strategy}

Provided the background on particle filters, tempering and SMC samplers, we are now ready to assemble our proposed solution. We will first propose our novel tempering idea suited for learning parameters $\theta$ in models on the form~\eqref{eq:ssm}, and then explore how the tempering pace can be automatically adapted to the problem. Thereafter we will provide an overview of the proposed algorithm, and detail some additional connections to existing literature.

\subsection{A tempering sequence for our problem}\label{sec:strategy:temp}

Our aim is to infer the posterior $p(\theta\mid y_{1:T})$
for the model~(\ref{eq:ssm}). The absence of measurement noise in~\eqref{eq:ssm:g} gives the likelihood estimate $z$ from the bootstrap particle filter a high variance (in particular for values of $\theta$ not explaining the data well), which is a problem when seeking $p(\theta\mid y_{1:T})$. We therefore suggest to introduce the modified model
\begin{subequations}\label{eq:ssm-T} 
	\begin{align}
	p(x_{t}\mid x_{1:t-1},\theta) & =f(x_{t}\mid x_{t-1},u_{t-1},\theta),\label{eq:ssm:f-T}\\
	y_{t} & =g(x_{t})+e_{t},\qquad\qquad{e_{t}\sim\mathcal{N}(0,\lambda_{p})}.\label{eq:ssm:g-T}
	\end{align}
\end{subequations}
This model has an artificial Gaussian\footnote{The choice of Gaussian noise is for convenience and clarity only. Other choices, for example heavy-tailed distributions, are also possible. The only requirement is that its density can be evaluated point-wise.} measurement noise with variance $\lambda_{p}$, and our original model (\ref{eq:ssm}) is recovered for $\lambda_{p}=0$. We denote the posterior distribution under this model as $p(\theta\mid y_{1:T},\lambda_{p})$, and the corresponding likelihood
$p(y_{1:T}\mid\theta,\lambda_{p})$.
Furthermore, we will define a decreasing sequence of $\lambda_{p}$, such that $\lambda_{P}=0$, and get a tempering sequence
\begin{equation}
\pi_{p}(\theta)=p(\theta\mid y_{1:T},\lambda_{p})\propto p(y_{1:T}\mid\theta,\lambda_{p}) p(\theta),\label{eq:tempp}
\end{equation}
which we have illustrated in Figure~\ref{fig:temp}. In this sequence, the target distribution (at $p=P)$ indeed becomes $\pi_{P}(\theta)$= $p(\theta\mid y_{1:T},\lambda_{P}=0)=p(\theta\mid y_{1:T})$, i.e., the posterior for $\theta$ in the original model~(\ref{eq:ssm}), the problem we study in this paper. Such a tempering sequence bears clear resemblance to the ABC methodology proposed in \cite{DDJ:2012}. However, to the best of the authors knowledge, such a tempering sequence has not previously been studied in the context of state-space models.

We will use the tempering sequence~\eqref{eq:tempp} in an SMC sampler. To this end, we need to be able to evaluate $\pi_{p}(\theta)$ up to proportionality. For this purpose, we propose to use the particle filter to estimate the likelihood $p(y_{1:T}\mid\theta)$ (and assume that the prior $p(\theta)$ can be evaluated). The algorithm will thus have a nested construction of SMC algorithms: the particle filter is used to generate likelihood estimates $z_p$ for
different values of~$\theta$ and~$\lambda_p$, and the SMC sampler is used to infer $\theta$ by keeping track of the samples $\psth{\theta^j}$ and deciding for which values of~$\theta$
to run the particle filter. However, to ease the presentation, we will throughout the rest of this section assume that we do have access to the likelihood $p(y_{1:T}\mid\theta,\lambda_{p})$ exactly. That is indeed the case if, for example, \eqref{eq:ssm} is a linear Gaussian state-space model or a finite discrete hidden Markov model, in which cases the Kalman filter \cite{Rugh:1993} or the forward-backward algorithm \cite{CMR:2005} would provide $p(y_{1:T}\mid\theta,\lambda_{p})$ exactly. We will later return (in Section~\ref{sec:full}) to the situation where we only have access to stochastic estimates $z_p$, and expand the algorithm with a few more details to ensure theoretical soundness also for the general (and practically interesting) case~\eqref{eq:ssm}.

\subsection{Automatically determining the tempering pace}

Choosing a good sequence $\{\lambda_{p}\}_{p=1}^{P}$ is fundamental
to the performance of the proposed algorithm. A sequence $\{\lambda_{p}\}_{p=0}^{P}$
that is decreasing too fast will lead to rapid changes in the landscape of
$\pi_p(\theta) = p(\theta\mid y_{1:T},\lambda_{p})$, obstructing the SMC sampler and adding to the variance of the final results. On the other hand, a sequence $\{\lambda_{p}\}_{p=0}^{P}$ that is decreasing too slowly will be a waste of computational power. To this end, we suggest to take inspiration from \citet{DDJ:2012}, where they tackle a somewhat similar problem with the same version of the SMC sampler. They argue that a good tempering sequence would yield an effective sample size (ESS, \cite{KLW:1994}) somewhat constant throughout the sequence $p=0,\dots,P$. The ESS is defined as
\begin{equation}
\text{ESS}\left(\{W^j_p\}_{j=1}^{N_\theta}\right) = \left(\sum_{j=1}^{N_\theta}\left(\frac{W_p^\ji}{\sum_{k=1}^{N_\theta}W_p^\ki}\right)^2\right)^{-1},
\end{equation}
where $W^j_p$ denotes the importance weight of sample $j$ from $\pi_p(\theta)$. The ESS takes values between $1$ and $N_\theta$, with the interpretation that inference based on the $N_\theta$ weighted samples is approximately equivalent to inference based on $\text{ESS}\left(\{W_p^j\}_{j=1}^{N_\theta}\right)$ equally weighted samples. Consequently, if the weight of a single sample dominates all the other, the ESS is 1, and if all samples have equal weights, the ESS is $N_\theta$. Intuitively, it is natural to expect that a smaller value of $\lambda_p$ gives a lower ESS, if the particles $x_t^n$ are fixed in the particle filter: the smaller $\lambda_p$, the fewer particles $x_t^n$ are likely to explain the measurement $y_t$, yielding a higher variance in the particle weights, and thus a low ESS. Furthermore, \citet{DDJ:2012} note that on their problem it is possible to solve the equation of setting $\lambda_{p}$ (note that $W_p^j$ depends on $\lambda_p$) such that
\begin{equation}
\text{ESS}\left(\{W_p^j\}_{j=1}^{N_\theta}\right) = \alpha N_\theta,\label{eq:esseq}
\end{equation}
where $\alpha$ is some user-chosen coefficient between $0$ and $1$. (A similar adaption can also be found in \cite{JSD+:2011}.) It turns out, perhaps a bit surprisingly, that it is in fact possible to solve~\eqref{eq:esseq} also in our case when $W_p^j$ depends on $z_p^j$ from the particle filter, which in turn depends on $\lambda_p$. 
We postpone the details to the subsequent section where we discuss the details of the inner particle filter algorithm which defines the estimate $z_p^j$.
By solving~\eqref{eq:esseq}, we obtain an automated way to determine the tempering pace `on the fly', i.e., automatically determining the value of each~$\lambda_p$ with the aim to achieve a constant `quality' (constant ESS) of the Monte Carlo approximation in runtime. 

\subsection{Termination}\label{sec:termination}

The variance of the estimates $z_p$ is likely to increase as $p$ increases and $\lambda_p$ approaches~$0$ (Section~\ref{sec:challenges}). The implications of an increased variance of $z_p$ will be that fewer of the proposed samples will be accepted in the Metropolis-Hastings step, and the overall performance of the SMC sampler will deteriorate. It may therefore be necessary to terminate the sampler prematurely (at, say, $\lambda_p = 0.01$ instead of the desired $\lambda_p=0$), and take the obtained samples as an approximate solution. One heuristic suggested by \cite{DDJ:2012}  for determining a suitable termination point is to monitor the rejection rate in the Metropolis-Hastings steps, and trigger a termination when it reaches a certain threshold. The effect of such a premature termination is analyzed (in a slightly different setting) by, e.g., \cite{DeanSJP:2014}.

\begin{algorithm}[tbh!]
	\KwOut{Samples $\psth{\theta^\ji}$ from $p(\theta\mid y_{1:T},\lambda_p)$.}
	\SetKwFunction{ESSp}{ESS}
	\SetKwFunction{Kp}{Metropolis-Hastings}
	Set $p\leftarrow 0$ and $\lambda_0$ large.\;
	Sample initial $\psth{{\theta}^\ji} \sim p(\theta\mid y_{1:T},\lambda_0)$ using, e.g., Metropolis-Hastings.\;
	\While{$\lambda$ not sufficiently small (Section~\ref{sec:termination})}{
		Update $p \leftarrow p+1$.\;
		Let $\omega^\ji \leftarrow p(y_{1:T}\mid \theta^\ji,\lambda_{p-1}) p(\theta^\ji)$.\;
		Find $\lambda_p$ such that \ESSp{$\psth{\omega^\ji},\{\widetilde\omega^\ji = p(y_{1:T}\mid \theta^\ji,\lambda_p)p(\theta^\ji) 
			\}_{^{j=1}}^{_{N_\theta}}$} $= \alpha\cdot N_\theta$.\;
		Let $\widetilde{\omega}^\ji \leftarrow p(\theta^\ji\mid y_{1:T},\lambda_p)$.\;
		Draw $\anc^\ji$ with $\Prb{\anc^\ji=k} \propto \frac{\widetilde{\omega}^\ki}{{\omega}^\ki}$.\;
		Sample $\theta^\ji \leftarrow $\Kp{$\lambda_p,\theta^\ji$}.\;
	}
	\setcounter{AlgoLine}{-1}
	\SetKwProg{myproc}{Function}{}{} 
	\myproc{\ESSp{$\psth{\omega^\ji},\psth{\widetilde{\omega}^\ji}$}}{
		Let $W^\ji \leftarrow \frac{\widetilde{\omega}^\ji}{\omega^\ji}$.\;
		\KwRet $\left(\sum_{j=1}^{N_\theta}\left(W^\ji/\sum_{k=1}^{N_\theta}W^\ki\right)^2\right)^{-1}$\;}
	\setcounter{AlgoLine}{-1}
	\SetKwProg{myproctt}{Function}{}{}
	\myproctt{\Kp{$\lambda_p,\theta^\ji$}}{
		Propose a new $\theta'~\sim q(\cdot\mid\theta^\ji)$.\;
		Sample $d \leftarrow \mathcal{U}_{[0,1]}$, i.e., uniformly on the interval $[0,1]$.\;
		\If{
			$d<\frac{p(y_{1:T}\mid \theta',\lambda_{p})p(\theta')}{p(y_{1:T}\mid \theta^\ji,\lambda_{p})p(\theta^\ji)}\frac{q(\theta^\ji\mid\theta')}{q(\theta'\mid\theta^\ji)}$}{
			Accept $\theta^\ji \leftarrow \theta'$.\;
		}
		\KwRet $\theta^\ji$
	}
	\setcounter{AlgoLine}{-1}
	\emph{(Lines with $j$ are for all $j = 1, \dots, N_\theta$).} \;
	\caption{Strategy for particle filter based learning of $\theta$ in~\eqref{eq:ssm}}
	\label{alg:simple}
\end{algorithm}

\subsection{Proposed algorithm -- preliminary version}

In Algorithm~\ref{alg:simple} we outline our proposed algorithm. Here, $q$ is the proposal in the Metropolis-Hastings sampler, proposing new values of the parameter $\theta$ which in a later stage are either accepted or rejected. 
We have in Algorithm~\ref{alg:simple} assumed that $p(y_{1:T} \mid \theta, \lambda_{p})$ can be evaluated exactly. However, in the general case of a nonlinear state-space model the particle filter has to be used, which results in an unbiased stochastic estimate  $z_p \approx p(y_{1:T}\mid \theta, \lambda_{p})$. We will address this fully in Section~\ref{sec:full}.

As mentioned earlier, a parallel to our problem with no measurement noise can be found in the literature under the heading approximate Bayesian computations (ABC,
\cite{BZB:2002}). In ABC, the idea is to simulate data $\widehat{y}(\theta)$ from a model and compare it to the recorded data $y$. ABC, however, is originally not formulated for state-space models, even though recent such contributions have been made \cite{Jasra:2015,DeanSJP:2014}. The introduction of an artificial measurement noise in our problem can be seen as an ABC-type of idea, but since the artificial measurement noise interacts with the particle filter (our analogy to simulate new data $\widehat{y}(\theta)$), our method does not qualify as a standard ABC solution.

Another closely related algorithm is the \SMCt algorithm \citep{CJP:2013}. \SMCt is also an SMC sampler using the particle filter to estimate the likelihood $z$, but it makes use of a data-tempered sequence (Section~\ref{sec:tempering}) instead of tempering based on artificial measurement noise~(\ref{eq:tempp}). For the problem of learning the parameters $\theta$ in~\eqref{eq:ssm}, the particle filter is likely to face troubles for small values of the measurement noise~$\lambda_{p}$. For our proposed algorithm, this can be handled by terminating the algorithm prematurely if necessary. Such a resort is not possible with a data-tempered sequence in \SMCt, since the problems with poor estimates $z$ from the particle filter would be faced already from the first step of a data-tempered sequence.

\section{Full algorithm and details}\label{sec:full}

In this section, we will first consider how to initialize the algorithm, and thereafter the details concerning the particle filter required for the adaptation of $\lambda_p$. Next, we present the proposed algorithm in detail and fully address the fact that the particle filter only provides stochastic estimates $z$, 
whereas Algorithm~\ref{alg:simple} requires that $p(y_{1:T}\mid\theta)$ can be evaluated exactly. The key is to consider the proposed algorithm to be sampling from an extended space explicitly encoding all randomness in the estimator $z$, 
and thereby reduce the problem to a standard SMC algorithm operating on an extended space.

\subsection{Initialization}

To initialize the SMC sampler properly, samples $\psth{\theta^\ji}$ from $p(\theta\mid y_{1:T},\lambda_0)$ are required. However, that distribution is typically not available to draw samples from directly. To this end, PMH \citep{SSM+:2017} (or \SMCt) can be used. Since $\lambda_0$ is user-chosen, we can choose it big enough such that  $p(y_{1:T}\mid\theta,\lambda_{0})$ has a broad support and we can obtain low-variance estimates $z_0$. However, in practice the use of Metropolis-Hastings inside the SMC sampler makes the algorithm somewhat `forgiving' with respect to initialization, and it may for practical purposes suffice to initialize the algorithm with samples $\psth{\theta^\ji}$ that are only approximate samples from $p(\theta\mid y_{1:T},\lambda_0)$ obtained using, e.g., some suboptimal optimization-based method.

\subsection{Re-visiting the particle filter}\label{sec:details:pf}

\begin{algorithm}[t]
	\KwIn{State space model $f(\farg\mid\farg,\theta)$, $g(\farg)$, $\lambda_p$,
		$p(x_{1})$, number of particles $N_x$, and data $y_{1:T}$.} 
	
	\KwOut{$x_{1:T},a_{2:T}$}
	
	Sample ${x}_{1}^{\ii}\sim p({x}_{1})$.\;
	
	Compute $w_{1}^{\ii}\leftarrow\mathcal{N}({y}_{1}|g({x}_{1}^{\ii}),\lambda_p)$.\;

	\For{$t=2$ \KwTo $T$}{
		Sample $\anc_{t}^{\ii}$ with $\Prb{\anc_{t}^{\ii}=j}\propto w_{t-1}^{\ji}$.\; 
		
		Sample ${x}_{t}^{\ii} \sim f(x_{t}|{x}_{t-1}^{\anc_{t}^{\ii}},\theta)$.\; 
		
		Compute $w_{t}^{\ii}\leftarrow\mathcal{N}({y}_{t}|g({x}_{t}^{\ii}),\lambda_p)$.\; 
		
	}
	
	\setcounter{AlgoLine}{-1}
	
	All operations are for $n=1, \dots, N_x$.\;
	\caption{Bootstrap particle filter}
	\label{alg:pf} 
\end{algorithm}

We have so far not fully justified the use of the particle filter inside the proposed algorithm. The particle filter provides a \emph{stochastic} estimate $z_p$ of $p(y_{1:T}\mid \theta, \lambda_p)$, and the $\lambda_p$-adaptation requires that we can solve \eqref{eq:esseq}, $\text{ESS}\left(\psth{W_p^j}\right) = \alpha N_\theta$, where $W_p^j$ depends on the ratio between
$z_p$ 
and
$z_{p-1}$, 
in turn depending on $\lambda_p$ and $\lambda_{p-1}$, respectively. Both estimates, $z_p$ and $z_{p-1}$, are stochastic, which seems not to allow for a well-defined numerical solution to~\eqref{eq:esseq}. This also implies that the weights $W^\ji$ in the SMC sampler are random themselves. The latter problem of stochastic weights within SMC is, however, already studied in the literature \cite{FPR+:2010}, whereas solving \eqref{eq:esseq} is novel in this work.

The key point for solving \eqref{eq:esseq} in our context with particle filters, and also to theoretically justify the random weights, is to consider the outcome of the particle filter (Algorithm~\ref{alg:pf}) to be all its internal random variables, 
$\psx{x^n_{1:T}, a^n_{2:T}}$, rather than only $z$. By doing so, we can explicitly handle all randomness in the particle filter, and understand our proposed algorithm as a standard algorithm on the non-standard extended space $\Theta\times\mathsf{X}^{N_xT}\times \mathsf{A}^{N_x(T-1)}$ (instead of only $\theta$), where $\mathsf{X}$ is the space in which $x_t$ lives, and similar for $\mathsf{A}$ and $a_t$. We will come back to this formalism, but let us first give a more intuitive view on the construction.

In solving \eqref{eq:esseq}, we would like to run the particle filter once (using $\lambda_{p-1}$), and afterwards decide on a $\lambda_p$ such that \eqref{eq:esseq} is fulfilled.
The random variables in the particle filter, 
$\psx{x_{1:T}^n,a_{2:T}^n}$,
are drawn with a certain distribution determined by the particle filter (Algorithm~\ref{alg:pf}) and $\lambda_p$. That is, if we were given samples $\psx{x_{1:T}^n,a_{2:T}^n}$
, we could compute the probability (density) of
$\psx{x_{1:T}^n,a_{2:T}^n}$
to be drawn by the particle filter. In particular, by inspection of Algorithm~\ref{alg:pf}, we realize that \emph{if} the ancestor variables 
$\psx{a_{2:T}^n}$
were fixed, $\lambda_p$ would not affect 
$\psx{x_{1:T}^n}$, but only the computation of $z$. Thus, if we run a particle filter with a measurement noise model with variance $\lambda_{p-1}$ and save 
$\psx{x_{1:T}^n,a_{2:T}^n}$,
we may afterwards compute the probability (density) of the resampling (i.e., the draw of
$\psx{a_{2:T}^n}$) to have happened had it been run with a measurement noise model with variance $\lambda_{p}$ instead\footnote{For this answer not to be exactly 0 forbiddingly often, multinomial resampling has to be used.}. This turns out to be enough for evaluating $\text{ESS}\left(\{W_p^j\}_{j=1}^{N_\theta}\right)$ conditionally on $\psx{x_{1:T}^n,a_{2:T}^n}$, which can be used to solve~\eqref{eq:esseq} using a numerical search, such as a bisection method.

This idea bears clear resemblances to the work by \cite{LeGland:2007}, but is not identical. Whereas \cite{LeGland:2007} considers fixed resampling weights across different models (in our context different $\lambda_p$), the resampling weights are not fixed in our approach, but changes with $\lambda_p$. 

A useful perspective is to understand our idea as importance sampling of $\psx{x_{1:T}^n,a_{2:T}^n}$, using a particle filter with $\lambda_{p-1}$ as proposal and a particle filter with $\lambda_{p}$ as target.

\begin{algorithm}[h!]
	\KwOut{Samples $\psth{\theta^\ji}$ from $p(\theta\mid y_{1:T},\lambda)$.}
	\SetKwFunction{ESSp}{ESS}
	\SetKwFunction{Wp}{w}
	\SetKwFunction{Kp}{Particle Metropolis-Hastings}
	Set $p\leftarrow0$ and $\lambda_0$ large.\;
	Sample initial $\psth{{\theta}^\ji} \sim p(\theta\mid y_{1:T},\lambda_0)$ using, e.g., particle Metropolis-Hastings.\;
	Run a particle filter with $\lambda_0$ for each $\theta^\ji$, and save $\pfv^\ji \triangleq \psx{x_{1:T}^\ii, a_{2:T}^\ii}$.\;
	\While{$\lambda_p$ not sufficiently small (Section~\ref{sec:termination})}{
		Update $p \leftarrow p+1$.\;
		Let $\omega^\ji \leftarrow $ \Wp{$\lambda_{p-1},\theta^\ji,\pfv^\ji$}.\;
		Find $\lambda_p$ such that \ESSp{$\psth{\omega^\ji},\{$\Wp{${\lambda_p},\theta^\ji,\pfv^\ji$}$\}_{^{j=1}}^{_{N_\theta}}$} $= \alpha\cdot N_\theta$.\;
		Let $\widetilde{\omega}^\ji \leftarrow $ \Wp{${\lambda_p},\theta^\ji,\pfv^\ji$}.\;
		Resample the $(\theta,\pfv)$-particles using weights $\propto \frac{\widetilde{\omega}^\ki}{{\omega}^\ki}$\;
		Sample $\{\theta^\ji,\pfv^\ji\} \leftarrow $\Kp{$\lambda_p,\theta^\ji,\pfv^\ji$}.\;
	}
	\setcounter{AlgoLine}{0}
	\SetKwProg{myproct}{Function}{}{}
	\myproct{\Wp{$\lambda_p,\theta^\ji,\pfv^\ji$}}{
		Let ${w}_{t-1}^\ii \leftarrow g({y}_t|{x}_t^\ii, {\lambda_p})$\;
		\KwRet $p(\theta^\ji)\left(\prod_{t=1}^{T}\sum_{n=1}^{N_x} {w}_{t-1}^\ii\right)\left(\prod_{t=1}^{T-1}\prod_{n=1}^{N_x} \Prb{a_{t+1}^\ii\mid \psx{{w}_{t}^\ii}}\right)$\;}
	\setcounter{AlgoLine}{0}
	\SetKwProg{myproc}{Function}{}{} 
	\myproc{\ESSp{$\psth{\omega^\ji},\psth{\widetilde{\omega}^\ji}$}}{
		Let $W^\ji \leftarrow \frac{\widetilde{\omega}^\ji}{\omega^\ji}$ for every $j$\;
		\KwRet $\left(\sum_{j=1}^{N_\theta}\left(W^\ji/\sum_{k=1}^{N_\theta}W^\ki\right)^2\right)^{-1}$\;}
	\setcounter{AlgoLine}{0}
	\SetKwProg{myproctt}{Function}{}{}
	\myproctt{\Kp{$\lambda_p,\theta^\ji,\pfv^\ji$}}{
		Let $z_p^j \leftarrow \prod_{t=1}^{T}\sum_{n=1}^{N_x} g(y_t\mid x_{t}^i, \lambda_p)$ (with $x_t^i$ from $\pfv^\ji$)\;
		Propose a new $\theta'~\sim q(\cdot\mid\theta^\ji)$ \;
		Run a particle filter with $\lambda_p$ and $\theta'$ and save $\pfv'$\;
		Let $z_p' \leftarrow \prod_{t=1}^{T}\sum_{n=1}^{N_x} g(y_t\mid x_{t}^i, \lambda_p)$ (with $x_t^i$ from $\pfv'$)\;
		Sample $d \leftarrow \mathcal{U}_{[0,1]}$, i.e., uniformly on the interval $[0,1]$.\;
		\If{$d<\frac{z_p' p(\theta')}{z_p^\ji p(\theta^\ji)}\frac{q(\theta^\ji\mid\theta')}{q(\theta'\mid\theta^\ji)}$}{
			Update $\theta^\ji \leftarrow \theta'$, $\pfv^\ji \leftarrow \pfv'$\;
		}
		\KwRet $\theta^\ji, \pfv^\ji$
	}
	\caption{Particle-filter based learning of $\theta$ in~\eqref{eq:ssm}}
	\label{alg:full}
\end{algorithm}

We summarize our proposed method in Algorithm~\ref{alg:full}. Continuing the extended space motivation,  Algorithm~\ref{alg:full} can in its most compact form be seen as a standard SMC sampler on the extended space $\Theta\times\mathsf{X}^{N_xT}\times \mathsf{A}^{N_x(T-1)}$ with target distribution at iteration $p$
\begin{multline}
p(\theta,\psx{x_{1:T},a_{2:T}}\mid y_{1:T},\lambda_p) \propto \\
p(\theta)
\left(\prod_{t=1}^{T}\sum_{n=1}^{N_{x}}g(y_{t}\mid x_{t}^{n},\lambda_p)\right)
\Bigg(\prod_{t=1}^{T-1}\prod_{n=1}^{N_{x}}\underbrace{\Prb{a_{t+1}^{\ii}\mid\psx{{w}_{t}^{\ii}}}}_{=\frac{g(y_{t}\mid x_{t}^{a_{t+1}^{n}},\lambda_p)}{\sum_{n=1}^{N_{x}}g(y_{t}\mid x_{t}^{n},\lambda_p)}}\Bigg)
\left(\prod_{t=1}^{T-1}\prod_{n=1}^{N_{x}}f(x_{t+1}^n\mid x_{t}^{a_{t+1}^n},\theta)\right).\label{eq:algtarg}
\end{multline}
From this, Algorithm~\ref{alg:full}, and in particular the particle filter (Algorithm~\ref{alg:pf}) as well as the weighting function \verb|w| in Algorithm~\ref{alg:full}, can be derived.

The previous paragraph can be understood as follows. First of all, the particle filter algorithm itself contains random elements. If we consider all randomness in the particle filter explicitly as random variables, i.e., consider
$\psx{x_{1:T}^n,a_{2:T}^n}$
and not just $z$, Algorithm~\ref{alg:full} is a standard SMC sampler \cite{DDJ:2006} for the distribution~\eqref{eq:algtarg}. This implies that available theoretical guarantees and convergence results (e.g., \cite{DDJ:2006,Chopin:2004,DelMoral:2004}) apply also to our construction when the $\lambda_p$ sequence is fixed. When $\lambda_p$ is selected adaptively these results do not readily apply, but \cite{BJK+:2016} have established convergence results for adaptive SMC algorithms in a related setting, and these results could possibly be extended to the adaptive scheme proposed in this article. Note also that no practical problems caused by the proposed adaptation have been encountered in the numerical examples.

\section{Numerical experiments}

In this section, we provide three numerical experiments illustrating and evaluating the proposed method from various perspectives. First, we start with a simple numerical example with a linear state-space model subject to Gaussian noise (implicitly introduced by Figure~\ref{fig:temp}) to illustrate the main ideas  presented by Algorithm~\ref{alg:simple}. We then consider a more challenging nonlinear example, where we compare our proposed method to the PMH algorithm~\cite{ADH:2010} and \SMCt~\cite{CJP:2013}, as well as a study on the influence of $T$. Finally we consider the challenging Wiener-Hammerstein benchmark problem~\cite{SN:2015}. The code for the examples is available via the first author's homepage.

\subsection{Toy example}\label{sec:numex:1}

\begin{figure}
	\setlength{\figureheight}{.25\linewidth}
	\setlength{\figurewidth}{.7\linewidth}
	\pgfplotsset{
		axis on top,
		label style={font=\scriptsize},
		legend style={inner xsep=1pt,inner ysep=0.5pt,nodes={inner sep=1pt,text depth=0.1em},font=\scriptsize},
		tick label style={font=\scriptsize},
		every axis/.append style={%
			scaled x ticks=false,
			scaled y ticks=false}
	}
	\centering
	\footnotesize
	\input{linex.tex}
	\caption{Result from applying Algorithm~\ref{alg:simple} to $T=200$ data points from the model~\eqref{eq:ex:lin}. The upper panels show how the marginals of the samples contracts as $\lambda_p\to0$ (cf. Figure~\ref{fig:temp}), and the lower panel shows the sequence $\{\lambda_p\}_{p=1}^P$ automatically determined by our algorithm in an adaptive manner.}
	\label{fig:linex}
\end{figure}
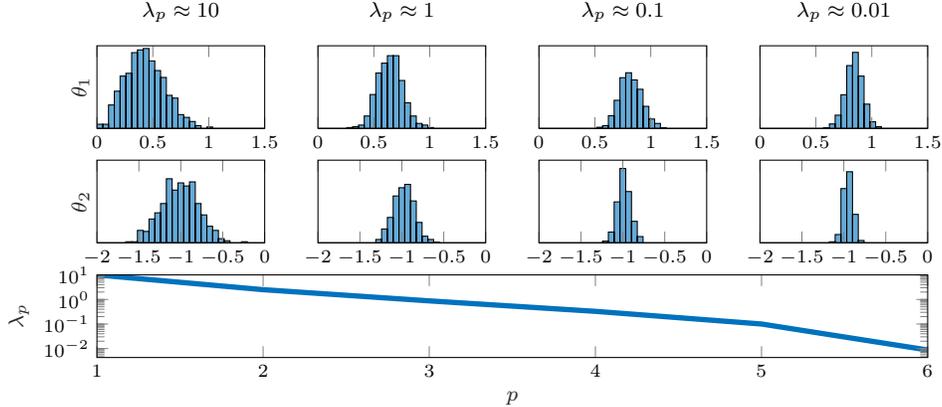

We consider the linear state-space model on the form
\begin{subequations}\label{eq:ex:lin}
	\begin{alignat}{3}
	x_{t+1} &= \begin{bmatrix}1 & \theta_1\\ 0 & 0.1\end{bmatrix}x_t + \begin{bmatrix}\theta_2\\ 0\end{bmatrix}u_t + v_t, \qquad & v_t&\sim \mathcal{N}\left(0,I_2\right),\\
	y_t &= \begin{bmatrix}1 & 0\end{bmatrix}x_t,
	\end{alignat}
\end{subequations}
where $\theta \triangleq \{\theta_1, \theta_2\}$ are the unknown parameters (true values: $\theta_1 = 0.8$, $\theta_2 = -1$) and $I_2$ denotes the identity matrix of dimension~$2$. This model was used to produce Figure~\ref{fig:temp}, where the propagation of samples $\psth{\theta^j}$ was illustrated. Since this model is linear and Gaussian, the  computation of the likelihood $p(y_{1:T}\mid\theta,\lambda_{p})$ can be done exactly\footnote{The choice of a linear and Gaussian model also made it  possible to exactly plot the contours in Figure~\ref{fig:temp}.} and no particle filter (with its potential problem due to small measurement noise variance) is needed. Thus, Algorithm~\ref{alg:simple} can be applied directly, by using the Kalman filter to exactly compute $p(y_{1:T}\mid\theta,\lambda_{p})$.  We now demonstrate Algorithm~\ref{alg:simple} by applying it to $T=200$ data points simulated from~\eqref{eq:ex:lin}. The artificial measurement noise $\lambda_p$ is automatically adapted such that ESS $\approx 0.5N_\theta$ at each step. The priors for both parameters are taken to be uniform on  $[0,2.5]\times[0,2.5]$. The resulting (marginal) posteriors are summarized in Figure~\ref{fig:linex}, which shows that the automatic tempering seems to work as expected. Figure~\ref{fig:temp} shows the true (joint) posteriors, and we can indeed confirm that their marginals resembles Figure~\ref{fig:linex}.

The main motivation behind our work was indeed to overcome the computational difficulties for the particle filter when the variance of the measurement noise is very small. However, for probabilistic learning of $\theta$ also in linear Gaussian models where the exact Kalman filter can be applied, sampling methods can still be useful for learning $\theta$, see, e.g, \cite{NH:2010,WSL+:2012} for the use of Metropolis-Hastings and Gibbs samplers, respectively. Our proposed tempering scheme for an SMC sampler thus presents yet another alternative for these models.

\subsection{A more challenging nonlinear example}\label{sec:numex:2}

\begin{figure}[t]
	\centering
	\setlength{\figureheight}{.3\linewidth}
	\setlength{\figurewidth}{.8\linewidth}
	\pgfplotsset{
		axis on top,
		label style={font=\scriptsize},
		legend style={inner xsep=1pt,inner ysep=0.5pt,nodes={inner sep=1pt,text depth=0.1em},font=\scriptsize},
		minor y tick num=1,
		minor x tick num=1,
		tick label style={font=\scriptsize},
		ylabel style={align=center}, 
		every axis/.append style={%
			scaled x ticks=false,
			scaled y ticks=false}
	}
	\centering
	\footnotesize
	\input{nepost_ny.tex}
	\caption{Posterior samples from the problem in Section~\ref{sec:numex:2}. The probability of acceptance (empirically $0.026\%$) in the Metropolis-Hastings mechanism is very low due to the model~\eqref{eq:numex:2:ssm} with highly informative observations, which gives a high variance in the estimates $z$. Neither PMH nor \SMCt therefore explore the posterior well, whereas the proposed method shows a better result due to the proposed tempering scheme, which adds an artificial measurement noise to the model giving it computational advantages with less variance in $z_p$.}
	\label{fig:nepost}
\end{figure}
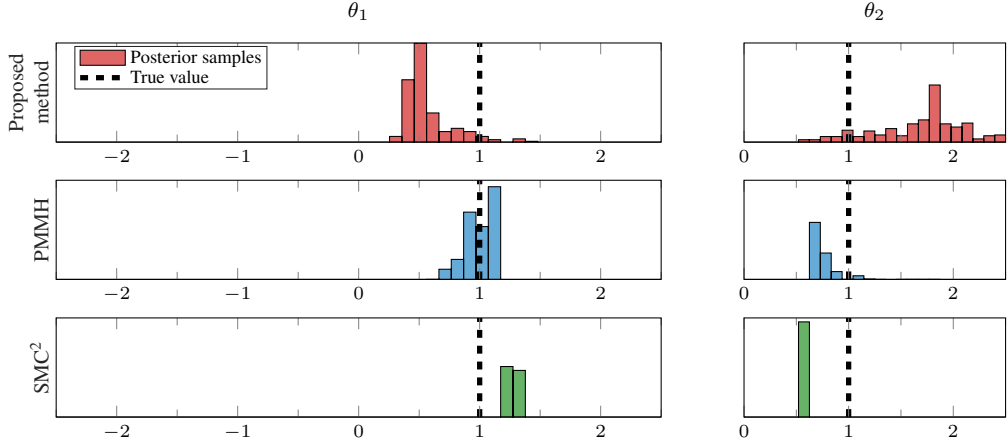

We now consider the following state-space model
\begin{subequations}\label{eq:numex:2:ssm}
	\begin{alignat}{2}
	x_{t+1} &= \text{atan}(x_t) + \theta_1u_t + v_t, &\qquad v_t&\sim \mathcal{N}(0,1),\\
	y_t &= |x_t| + \theta_1\theta_2 + e_t, &e_t&\sim \mathcal{N}(0,10^{-2}).\label{eq:numex:2:ssm:g}
	\end{alignat}
\end{subequations}
This model, with a $1$-dimensional state space, has as an exogenous input $u_t$, a significant amount of process noise $v_t$ and an almost negligible measurement noise $e_t$. From this model, $T=300$ data points were simulated and the two unknown parameters $\theta = \{\theta_1, \theta_2\}$ are to be learned from the measured data $\{y_{1:T}, u_{1:T}\}$ with uniform priors. The input $u_{1:T}$ is taken as a realization of a white noise random process.

The relatively short data record together with the presence of $\theta_2$ only in the product $\theta_1\theta_2$ in~\eqref{eq:numex:2:ssm:g} suggest there is a certain amount of uncertainty present in the problem, which we expect to be reflected in the posterior. 
However, the highly informative observations makes this a rather challenging problem for the standard methods.

We apply our proposed method, and compare it to PMH \cite{DDJ:2006} and \SMCt \cite{CJP:2013} on this problem. In all algorithms, we use the bootstrap particle filter (Algorithm~\ref{alg:pf}) with $N_x=300$, and a simple random walk proposal. Furthermore, we let $N_{\theta} = 300$, $K=40$ and $\alpha = 0.3$ in Algorithm~\ref{alg:full}, as well as their counterparts in \SMCt, and we run PMH until $100\,000$ samples are obtained. We use the same Metropolis-Hastings proposal in all algorithms. For our proposed algorithm, we adopt a similar heuristic as \cite{DDJ:2012} and terminate the tempering once the acceptance rate in the Metropolis-Hastings procedure goes below $5\%$.

The obtained posterior samples are shown in Figure~\ref{fig:nepost}. The mixing of PMH is rather poor (the acceptance rate was recorded as $0.026\%$), and it has consequently not managed to explore the posterior as well as our proposed method. A similar problem occurs for \SMCt, which performs even worse on this problem. Since the tempering in our proposed method, however, follows a sequence of decreasing artificial measurement noise, which terminates once the mixing becomes too bad, it does not suffer from the same problem. 

The settings of PMH can indeed be optimized by using more clever proposals than random walks (see, e.g., \cite{DS:2015} for an overview) and methods for reducing the variance of $z$ (such as adapted or bridging particle filter \cite{DM:2015,PS:1999,DJL+:2015}). Such adaption would indeed push the performance further. 
However, the adaption could be applied to all three methods, and such tuning is therefore not crucial in a relative comparison between the methods.

\subsection{Evaluating the performance with growing $T$}\label{sec:numex:3}

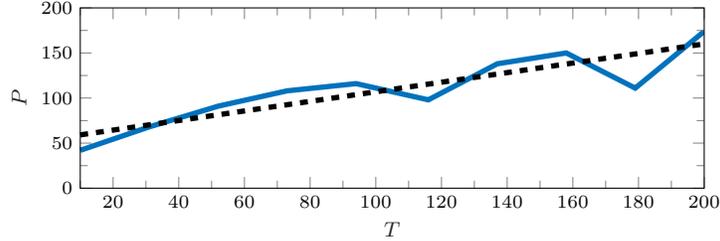
\begin{figure}[t]
	\centering
	\setlength{\figureheight}{.15\linewidth}
	\setlength{\figurewidth}{.5\linewidth}
	\pgfplotsset{
		axis on top,
		label style={font=\scriptsize},
		legend style={inner xsep=1pt,inner ysep=0.5pt,nodes={inner sep=1pt,text depth=0.1em},font=\scriptsize},
		minor y tick num=1,
		minor x tick num=1,
		tick label style={font=\scriptsize},
		ylabel style={align=center}, 
		every axis/.append style={%
			scaled x ticks=false,
			scaled y ticks=false}
	}
	\centering
	\footnotesize
	\input{ex3.tex}
	\caption{The number of steps (solid blue) required in Example~\ref{sec:numex:3} to transition from the initial $\lambda_{0}=1$ to the true final $\lambda_P=0.05$ when different number of data points $T$ are included in the data set, and $\alpha$ is fixed at $0.4$. As expected the number of steps, $P$, required grows with the number of data points included, $T$, seemingly in a rather linear way (dashed black).}
	\label{fig:ex3}
\end{figure}

As discussed in Section~\ref{sec:details:pf}, the proposed method can essentially be understood as an importance sampler producing $\{x_{1:T}^n,a_{2:T}^n\}_{n=1}^{N_x}$. Because of this, we could expect the method to be less efficient as the number of measurements $T$ grows, since $T$ is one of the dimensions in the importance sampling space. We study this effect with a similar state-space model as above, and record how many steps $P$ that are required for the artificial output noise to transition from the initial starting point $\lambda_{0}=1$ to the true $\lambda_{P}=0.05$, when $\alpha$ is set to $0.4$, and different number of data points $T$ are included in the data set. The results are shown in Figure~\ref{fig:ex3} and suggest a linear growth in the number of steps $P$ as $T$ grows. In combination with a computational load growing linearly with $T$ for the particle filter, the total computational load is $\propto T^2$ for $N_x$ constant\footnote{For optimal performance, possibly also $N_x$ should be scaled with $T$, this is a question for further research.}.

\subsection{The Wiener-Hammerstein benchmark with process noise}

The Wiener-Hammerstein benchmark \cite{SN:2015} is a recent benchmark problem for system identification which is a particular special case of the model~\eqref{eq:ssm}. This problem has also served as the motivating problem for us to propose this method. The benchmark is implemented as an electronic circuit, and the challenge is to use recorded data from the system to estimate a model which is able to imitate the behavior of the electric circuit well. The system can be described as a Wiener-Hammerstein system, i.e., a linear dynamical system, a static nonlinearity, and then another linear dynamical system in series. This is by now a fairly well-studied model, see e.g. \cite{GiriB:2010,SchoukensSL:2009,BillingsF:1982,BershadCM:2001} for earlier work. There was also a relatively recent special section devoted to an earlier Wiener-Hammerstein benchmark problem in Control Engineering Practice in $2012$~\cite{HjalmarssonRR:2012}. The key difference is the significantly higher process noise level in this newly proposed benchmark.

The input to the system is a (known) signal entering into the first linear system. There is also an (unknown) colored process noise present, which enters directly into the nonlinearity. The measurements are of rather high quality, so there is very little measurement noise (when compared to the process noise) which makes the measurements highly informative. The system is summarized in Figure~\ref{fig:WHs}.

\tikzstyle{block} = [draw, rectangle, minimum height=2em, minimum width=2em]

\begin{figure}[h]
	\centering
	\begin{subfigure}{.6\linewidth}
		\centering
		\begin{tikzpicture}[auto, node distance=4em]
		\node [label=left:{$u_t$}] (u) {};
		\node [block, right of=u] (G) {$G$};
		\node [right of=G, draw, circle, minimum size=1em, inner sep=0pt] (sum) {$+$};
		\node [block, right of=sum] (f) {$f$};
		\node [block,right of=f] (S) {$S$};
		\node [right of=S,label=right:{$y_t$}] (y) {};
		\node [block,above of=G] (H) {$H$};
		\node [left of=H,label=left:{$v_t$}] (e) {};
		
		\draw [->] (u) -- (G);
		\draw [->] (G) -- (sum);
		\draw [->] (sum) -- (f);
		\draw [->] (f) -- (S);
		\draw [->] (S) -- (y);
		\draw [->] (H) -| (sum);
		\draw [->] (e) -- (H);
		\end{tikzpicture}
		\caption{A block diagram describing the system. The blocks with uppercase letters ($G$, $H$, $S$) are all linear dynamical systems, whereas the block with lowercase $f$ is a static nonlinearity. Further, $u_t$ is a known input signal, $v_t$ is (white) process noise and $y_t$ is the measured output.}
		\label{fig:WHblock}
	\end{subfigure}
	\hspace{1em}
	\begin{subfigure}{.35\linewidth}
		\centering
		\includegraphics[width=.8\linewidth]{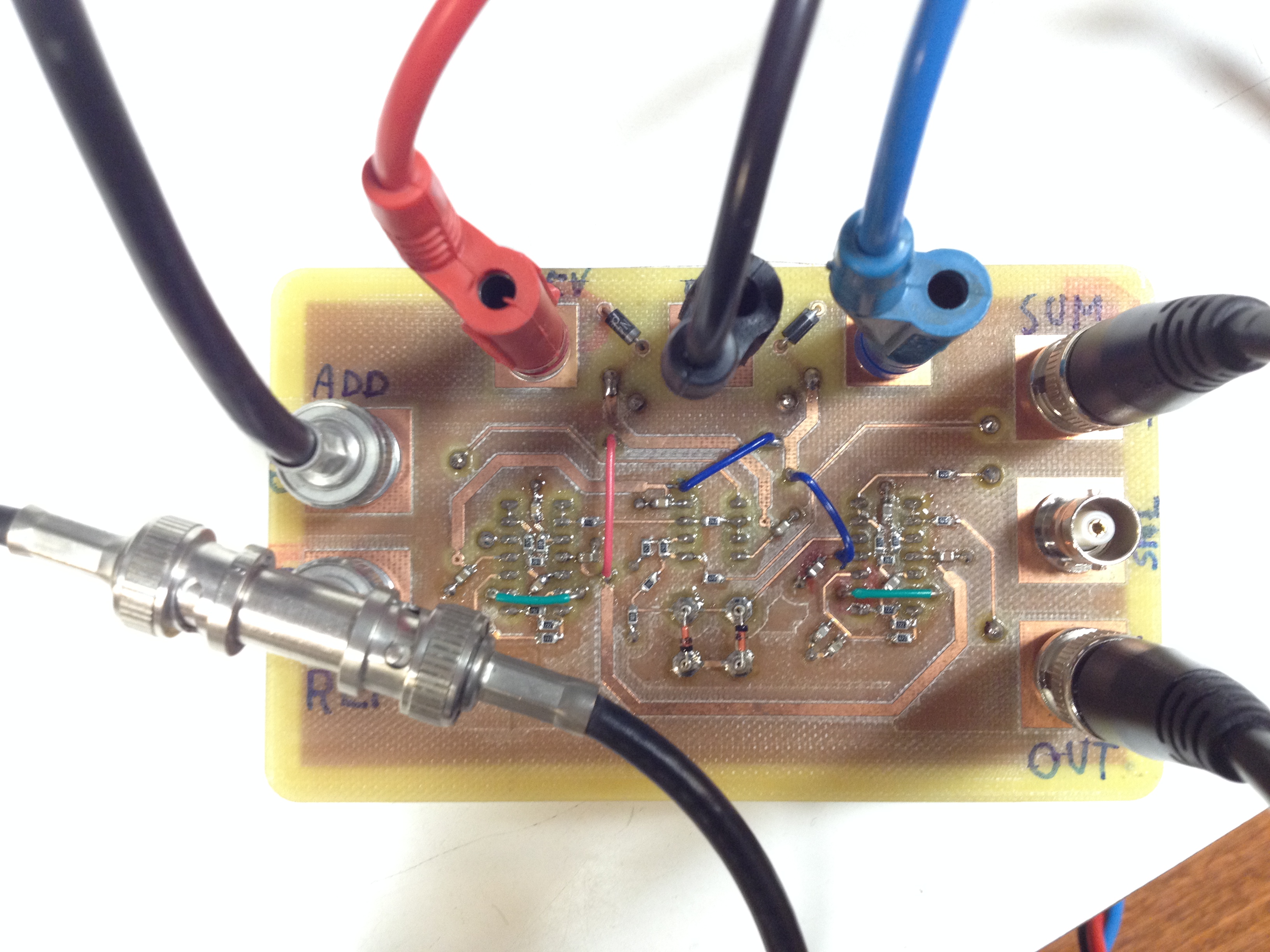}
		\caption{A picture of the electronic circuit}
		\label{fig:WHpic}
	\end{subfigure}
	\caption{The Wiener-Hammerstein benchmark system.}
	\label{fig:WHs}
\end{figure}
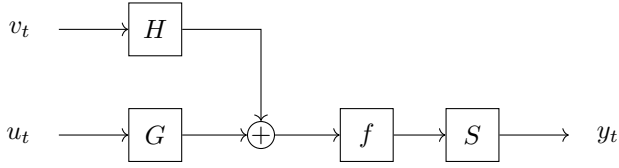
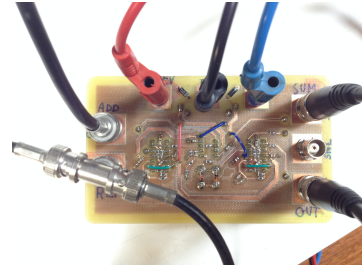

The structure of the Wiener-Hammerstein benchmark system can be brought into the state-space formalism as
\begin{subequations}
	\begin{align}
	\notag
	\begin{bmatrix}
	x_{t+1}^G \\~\\ x_{t+1}^H \\~\\ x_{t+1}^S
	\end{bmatrix}
	&=
	\begin{bmatrix}
	{A^G} & 0 & 0 \\~& & \\ 0 & {A^H} & 0 \\~&  & \\ 0 & 0 & {A^S}
	\end{bmatrix}
	\begin{bmatrix}
	x_{t}^G \\~\\ x_{t}^H \\~\\ x_{t}^S
	\end{bmatrix}
	+
	\begin{bmatrix}
	{B^G} \\~\\ 0 \\~\\ 0
	\end{bmatrix}
	u_t
	+
	\begin{bmatrix}
	0 \\~\\ {B^H} \\~\\ 0
	\end{bmatrix}
	v_t\\
	&+
	\begin{bmatrix}
	0 \\~\\ 0 \\~\\ B^S
	\end{bmatrix}
	\sum_{m=1}^M {c^{(m)}}\phi^{(m)}(C^Gx_t^G + C^Hx_t^H + D^Gu_t),\\
	y_t &= {C^S}x_t^S + {D^S}v_t^S,
	\end{align}
\end{subequations}
where all $A$s are $3\times3$-matrices, $B$s are $3\times1$-matrices, $C$s are $1\times3$-matrices, $D$s are scalars, $\{\phi^{(m)}\}$ is a Fourier basis function expansion (truncated at $M=10$), and $v_t$ is a zero-mean scalar-valued white Gaussian process noise with unknown variance. Adjusting for the overparametrization of the linear state-space model, the effective number of unknown parameters is $30$. For learning the parameters, a data set with $T=8192$ samples and $u_t$ a faded multisine input\footnote{Available as \texttt{WH\_MultisineFadeOut} at \texttt{http://homepages.vub.ac.be/{\textasciitilde}mschouke/benchmarkWienerHammerstein.html}.}
was used. We applied Algorithm~\ref{alg:full} with $N_\theta=50$ and $K=10$. For initialization purposes, an approximate model was found essentially using the ideas by~\cite{PLS+:2010} (which is computationally lighter, but cannot fully handle the presence of process noise, on the contrary to Algorithm~\ref{alg:full}).

\begin{figure}[t]
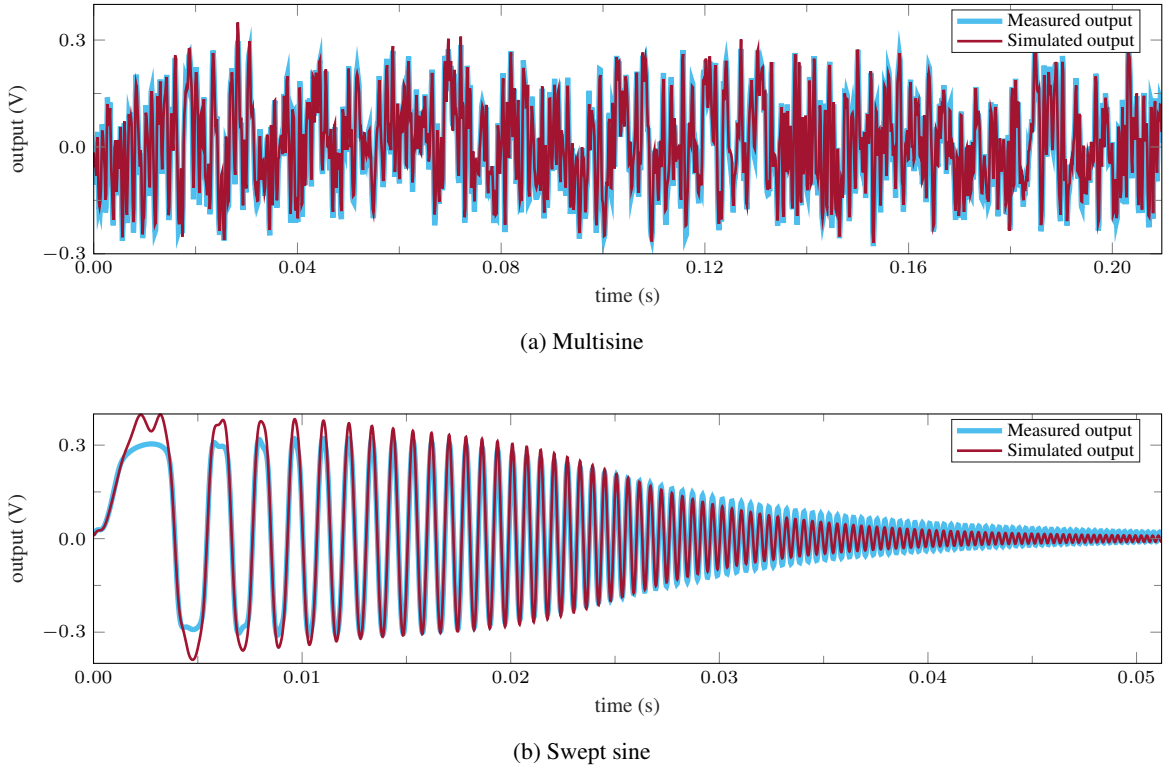

	\centering
	\begin{subfigure}[t]{\linewidth}
		\setlength{\figureheight}{.2\linewidth}
		\setlength{\figurewidth}{.9\linewidth}
		\pgfplotsset{
			axis on top,
			label style={font=\scriptsize},
			legend style={inner xsep=1pt,inner ysep=0.5pt,nodes={inner sep=1pt,text depth=0.1em},font=\scriptsize},
			minor y tick num=1,
			minor x tick num=1,
			tick label style={font=\scriptsize},
			every axis/.append style={%
				scaled x ticks=false,
				scaled y ticks=false,
				yticklabel style={/pgf/number format/.cd,fixed,fixed zerofill,precision=1},
				xticklabel style={/pgf/number format/.cd,fixed,fixed zerofill,precision=2}}
		}
		\centering
		\footnotesize
		\input{WHms.tex}
		\caption{Multisine}
		\label{fig:WHms}
	\end{subfigure}
	
	\vspace{20pt}
	
	\begin{subfigure}[t]{\linewidth}
		\setlength{\figureheight}{.2\linewidth}
		\setlength{\figurewidth}{.9\linewidth}
		\pgfplotsset{
			axis on top,
			label style={font=\scriptsize},
			legend style={inner xsep=1pt,inner ysep=0.5pt,nodes={inner sep=1pt,text depth=0.1em},font=\scriptsize},
			minor y tick num=1,
			minor x tick num=1,
			tick label style={font=\scriptsize},
			every axis/.append style={%
				scaled x ticks=false,
				scaled y ticks=false,
				yticklabel style={/pgf/number format/.cd,fixed,fixed zerofill,precision=1},
				xticklabel style={/pgf/number format/.cd,fixed,fixed zerofill,precision=2}}
		}
		\centering
		\footnotesize
		\input{WHss.tex}
		\caption{Swept sine}
		\label{fig:WHss}
	\end{subfigure}
	\caption{Simulated output from the model (red line) versus the recorded output (blue line) for the Wiener-Hammerstein benchmark. The test data sets are recorded with no process noise, as opposed to the training data sets that were used for learning the model.}
	\label{fig:WH}
\end{figure}

The obtained results are presented in Table~\ref{tab:WH} and Figure~\ref{fig:WH}, where they are reported according to the benchmark instructions \cite{SN:2015}, i.e., the simulation error for two test data sets measured on the system with no process noise present and a swept sine and a multisine as input, respectively. For reference, the performance of the model used to initialize Algorithm~\ref{alg:full} is also included. The results reported were obtained within a few hours on a standard personal computer.

The essentially non-existing measurement noise makes PMH (as well as \SMCt) incompatible with this problem.

\begin{table}
	\centering
	\caption{Wiener-Hammerstein benchmark: The root mean square error (RMSE) of the simulation error on the provided test data sets.}
	\begin{tabular}{c c c}
		& RMSE of proposed method & RMSE of initial model \\ \hline
		Swept sine & 0.014 & 0.039 \\
		Multisine & 0.015 & 0.038 \\
	\end{tabular}
	\label{tab:WH}
\end{table}

\section{Discussion}

We have proposed an algorithm for probabilistic learning of unknown parameters in models of the structure~\eqref{eq:ssm}, i.e., state-space models with highly informative observations. Our proposed algorithm can be understood as either an ABC-inspired methodology \cite{DeanSJP:2014,DDJ:2012}, or as an alternative tempering in an SMC sampler (akin to \SMCt \cite{CJP:2013}). Its theoretical justification follows from viewing it as a standard SMC sampler on an extended space, and well established theoretical guarantees are thus available.

The importance sampling perspective (Section~\ref{sec:details:pf}) raises the question of how well the proposed  adaptation of $\lambda_p$ scales with dimensionality, in particular the number of measurements $T$. This is partly investigated in Example~\ref{sec:numex:3} suggesting (at least) a computational load $\propto T^2$, but we also note that the method performs well in the benchmark example containing $T=8192$ samples. A more systematic numerical evaluation of the proposed method, which however is beyond the scope of this paper, would certainly be of interest.

For further research, connections with the idea of variational tempering \cite{MMA+:2016} could possibly also be of interest to explore. It is also not obvious that the ESS criterion~\eqref{eq:esseq} is the best criterion for deciding a well-performing tempering within the SMC sampler, and other alternatives could be studied and compared.

For optimal performance, our proposed method could be combined with methods for variance reduction of the estimate $z$, such as the adapted or bridging particle filter \cite{PS:1999,DM:2015,DJL+:2015}. The combination with such methods would indeed be interesting to explore further. However, 
while the use of such methods may indeed push the limits, for most cases they will not remove the fundamental problem.


\bibliographystyle{abbrvnat}
\bibliography{references}

\end{document}

%% file: linex.tex
%
%
\definecolor{mycolor1}{rgb}{0.00000,0.44700,0.74100}%
\begin{tikzpicture}

\begin{axis}[%
width=0.192\figurewidth,
height=0.265\figureheight,
at={(0\figurewidth,0.735\figureheight)},
scale only axis,
xmin=0,
xmax=1.5,
ymin=0,
ymax=60,
ytick={\empty},
ylabel style={font=\color{white!15!black}},
ylabel={$\theta_1$},
axis background/.style={fill=white},
title style={font=\bfseries},
title={$\lambda_p \approx 10$}
]
\addplot[ybar interval, fill=mycolor1, fill opacity=0.6, draw=black, area legend] table[row sep=crcr] {%
x	y\\
0	3\\
0.0517241379310345	3\\
0.103448275862069	17\\
0.155172413793103	27\\
0.206896551724138	38\\
0.258620689655172	40\\
0.310344827586207	55\\
0.362068965517241	57\\
0.413793103448276	58\\
0.46551724137931	49\\
0.517241379310345	42\\
0.568965517241379	37\\
0.620689655172414	24\\
0.672413793103448	22\\
0.724137931034483	10\\
0.775862068965517	8\\
0.827586206896552	5\\
0.879310344827586	2\\
0.931034482758621	0\\
0.982758620689655	1\\
1.03448275862069	0\\
1.08620689655172	0\\
1.13793103448276	0\\
1.18965517241379	0\\
1.24137931034483	0\\
1.29310344827586	0\\
1.3448275862069	0\\
1.39655172413793	0\\
1.44827586206897	0\\
1.5	0\\
};
\end{axis}

\begin{axis}[%
width=0.192\figurewidth,
height=0.265\figureheight,
at={(0\figurewidth,0.368\figureheight)},
scale only axis,
xmin=-2,
xmax=0,
ymin=0,
ymax=80,
ytick={\empty},
ylabel style={font=\color{white!15!black}},
ylabel={$\theta_2$},
axis background/.style={fill=white}
]
\addplot[ybar interval, fill=mycolor1, fill opacity=0.6, draw=black, area legend] table[row sep=crcr] {%
x	y\\
-2	0\\
-1.93103448275862	0\\
-1.86206896551724	0\\
-1.79310344827586	0\\
-1.72413793103448	0\\
-1.6551724137931	1\\
-1.58620689655172	1\\
-1.51724137931034	12\\
-1.44827586206897	11\\
-1.37931034482759	23\\
-1.31034482758621	29\\
-1.24137931034483	38\\
-1.17241379310345	62\\
-1.10344827586207	51\\
-1.03448275862069	58\\
-0.96551724137931	55\\
-0.896551724137931	59\\
-0.827586206896552	38\\
-0.758620689655172	27\\
-0.689655172413793	16\\
-0.620689655172414	12\\
-0.551724137931034	4\\
-0.482758620689655	2\\
-0.413793103448276	0\\
-0.344827586206897	0\\
-0.275862068965517	1\\
-0.206896551724138	0\\
-0.137931034482759	0\\
-0.0689655172413792	0\\
0	0\\
};
\end{axis}

\begin{axis}[%
width=0.192\figurewidth,
height=0.265\figureheight,
at={(0.253\figurewidth,0.735\figureheight)},
scale only axis,
xmin=0,
xmax=1.5,
ymin=0,
ymax=100,
ytick={\empty},
axis background/.style={fill=white},
title style={font=\bfseries},
title={$\lambda_p \approx 1$}
]
\addplot[ybar interval, fill=mycolor1, fill opacity=0.6, draw=black, area legend] table[row sep=crcr] {%
x	y\\
0	0\\
0.0517241379310345	0\\
0.103448275862069	0\\
0.155172413793103	0\\
0.206896551724138	0\\
0.258620689655172	1\\
0.310344827586207	2\\
0.362068965517241	4\\
0.413793103448276	15\\
0.46551724137931	41\\
0.517241379310345	67\\
0.568965517241379	79\\
0.620689655172414	88\\
0.672413793103448	88\\
0.724137931034483	58\\
0.775862068965517	31\\
0.827586206896552	15\\
0.879310344827586	6\\
0.931034482758621	4\\
0.982758620689655	1\\
1.03448275862069	0\\
1.08620689655172	0\\
1.13793103448276	0\\
1.18965517241379	0\\
1.24137931034483	0\\
1.29310344827586	0\\
1.3448275862069	0\\
1.39655172413793	0\\
1.44827586206897	0\\
1.5	0\\
};
\end{axis}

\begin{axis}[%
width=0.192\figurewidth,
height=0.265\figureheight,
at={(0.253\figurewidth,0.368\figureheight)},
scale only axis,
xmin=-2,
xmax=0,
ymin=0,
ymax=150,
ytick={\empty},
axis background/.style={fill=white}
]
\addplot[ybar interval, fill=mycolor1, fill opacity=0.6, draw=black, area legend] table[row sep=crcr] {%
x	y\\
-2	0\\
-1.93103448275862	0\\
-1.86206896551724	0\\
-1.79310344827586	0\\
-1.72413793103448	0\\
-1.6551724137931	0\\
-1.58620689655172	0\\
-1.51724137931034	0\\
-1.44827586206897	0\\
-1.37931034482759	0\\
-1.31034482758621	6\\
-1.24137931034483	25\\
-1.17241379310345	46\\
-1.10344827586207	85\\
-1.03448275862069	100\\
-0.96551724137931	106\\
-0.896551724137931	76\\
-0.827586206896552	37\\
-0.758620689655172	12\\
-0.689655172413793	6\\
-0.620689655172414	1\\
-0.551724137931034	0\\
-0.482758620689655	0\\
-0.413793103448276	0\\
-0.344827586206897	0\\
-0.275862068965517	0\\
-0.206896551724138	0\\
-0.137931034482759	0\\
-0.0689655172413792	0\\
0	0\\
};
\end{axis}

\begin{axis}[%
width=0.192\figurewidth,
height=0.265\figureheight,
at={(0.506\figurewidth,0.735\figureheight)},
scale only axis,
xmin=0,
xmax=1.5,
ymin=0,
ymax=150,
ytick={\empty},
axis background/.style={fill=white},
title style={font=\bfseries},
title={$\lambda_p \approx 0.1$}
]
\addplot[ybar interval, fill=mycolor1, fill opacity=0.6, draw=black, area legend] table[row sep=crcr] {%
x	y\\
0	0\\
0.0517241379310345	0\\
0.103448275862069	0\\
0.155172413793103	0\\
0.206896551724138	0\\
0.258620689655172	0\\
0.310344827586207	0\\
0.362068965517241	0\\
0.413793103448276	0\\
0.46551724137931	0\\
0.517241379310345	2\\
0.568965517241379	5\\
0.620689655172414	23\\
0.672413793103448	51\\
0.724137931034483	92\\
0.775862068965517	101\\
0.827586206896552	88\\
0.879310344827586	71\\
0.931034482758621	40\\
0.982758620689655	17\\
1.03448275862069	9\\
1.08620689655172	1\\
1.13793103448276	0\\
1.18965517241379	0\\
1.24137931034483	0\\
1.29310344827586	0\\
1.3448275862069	0\\
1.39655172413793	0\\
1.44827586206897	0\\
1.5	0\\
};
\end{axis}

\begin{axis}[%
width=0.192\figurewidth,
height=0.265\figureheight,
at={(0.506\figurewidth,0.368\figureheight)},
scale only axis,
xmin=-2,
xmax=0,
ymin=0,
ymax=200,
ytick={\empty},
axis background/.style={fill=white}
]
\addplot[ybar interval, fill=mycolor1, fill opacity=0.6, draw=black, area legend] table[row sep=crcr] {%
x	y\\
-2	0\\
-1.93103448275862	0\\
-1.86206896551724	0\\
-1.79310344827586	0\\
-1.72413793103448	0\\
-1.6551724137931	0\\
-1.58620689655172	0\\
-1.51724137931034	0\\
-1.44827586206897	0\\
-1.37931034482759	0\\
-1.31034482758621	0\\
-1.24137931034483	4\\
-1.17241379310345	26\\
-1.10344827586207	99\\
-1.03448275862069	180\\
-0.96551724137931	124\\
-0.896551724137931	52\\
-0.827586206896552	15\\
-0.758620689655172	0\\
-0.689655172413793	0\\
-0.620689655172414	0\\
-0.551724137931034	0\\
-0.482758620689655	0\\
-0.413793103448276	0\\
-0.344827586206897	0\\
-0.275862068965517	0\\
-0.206896551724138	0\\
-0.137931034482759	0\\
-0.0689655172413792	0\\
0	0\\
};
\end{axis}

\begin{axis}[%
width=0.192\figurewidth,
height=0.265\figureheight,
at={(0.759\figurewidth,0.735\figureheight)},
scale only axis,
xmin=0,
xmax=1.5,
ymin=0,
ymax=150,
ytick={\empty},
axis background/.style={fill=white},
title style={font=\bfseries},
title={$\lambda_p \approx 0.01$}
]
\addplot[ybar interval, fill=mycolor1, fill opacity=0.6, draw=black, area legend] table[row sep=crcr] {%
x	y\\
0	0\\
0.0517241379310345	0\\
0.103448275862069	0\\
0.155172413793103	0\\
0.206896551724138	0\\
0.258620689655172	0\\
0.310344827586207	0\\
0.362068965517241	0\\
0.413793103448276	0\\
0.46551724137931	0\\
0.517241379310345	0\\
0.568965517241379	1\\
0.620689655172414	8\\
0.672413793103448	24\\
0.724137931034483	61\\
0.775862068965517	114\\
0.827586206896552	138\\
0.879310344827586	95\\
0.931034482758621	44\\
0.982758620689655	12\\
1.03448275862069	3\\
1.08620689655172	0\\
1.13793103448276	0\\
1.18965517241379	0\\
1.24137931034483	0\\
1.29310344827586	0\\
1.3448275862069	0\\
1.39655172413793	0\\
1.44827586206897	0\\
1.5	0\\
};
\end{axis}

\begin{axis}[%
width=0.192\figurewidth,
height=0.265\figureheight,
at={(0.759\figurewidth,0.368\figureheight)},
scale only axis,
xmin=-2,
xmax=0,
ymin=0,
ymax=250,
ytick={\empty},
axis background/.style={fill=white}
]
\addplot[ybar interval, fill=mycolor1, fill opacity=0.6, draw=black, area legend] table[row sep=crcr] {%
x	y\\
-2	0\\
-1.93103448275862	0\\
-1.86206896551724	0\\
-1.79310344827586	0\\
-1.72413793103448	0\\
-1.6551724137931	0\\
-1.58620689655172	0\\
-1.51724137931034	0\\
-1.44827586206897	0\\
-1.37931034482759	0\\
-1.31034482758621	0\\
-1.24137931034483	0\\
-1.17241379310345	3\\
-1.10344827586207	25\\
-1.03448275862069	164\\
-0.96551724137931	215\\
-0.896551724137931	86\\
-0.827586206896552	7\\
-0.758620689655172	0\\
-0.689655172413793	0\\
-0.620689655172414	0\\
-0.551724137931034	0\\
-0.482758620689655	0\\
-0.413793103448276	0\\
-0.344827586206897	0\\
-0.275862068965517	0\\
-0.206896551724138	0\\
-0.137931034482759	0\\
-0.0689655172413792	0\\
0	0\\
};
\end{axis}

\begin{semilogyaxis}[%
width=0.951\figurewidth,
height=0.265\figureheight,
at={(0\figurewidth,0\figureheight)},
scale only axis,
xmin=1,
xmax=6,
xtick={0, 1, 2, 3, 4, 5, 6},
xlabel style={font=\color{white!15!black}},
xlabel={$p$},
ymin=0,
ymax=10,
ylabel style={font=\color{white!15!black}},
ylabel={$\lambda_p$},
axis background/.style={fill=white}
]
\addplot [color=mycolor1, line width=2.0pt, forget plot]
  table[row sep=crcr]{%
1	10\\
2	2.5\\
3	0.875\\
4	0.328125\\
5	0.1\\
6	0.00875\\
};
\end{semilogyaxis}
\end{tikzpicture}%

%% file: nepost_ny.tex
%
%
\definecolor{mycolor1}{rgb}{0.00000,0.44700,0.74100}%
\begin{tikzpicture}

\begin{axis}[%
width=0.606\figurewidth,
height=0.265\figureheight,
at={(0\figurewidth,0.735\figureheight)},
scale only axis,
xmin=-2.5,
xmax=2.5,
ymin=0,
ymax=4,
ytick={\empty},
ylabel style={font=\color{white!15!black}},
ylabel={Proposed\\method},
axis background/.style={fill=white},
title style={font=\bfseries},
title={$\theta_1$},
legend style={at={(0.03,0.97)}, anchor=north west, legend cell align=left, align=left, draw=white!15!black}
]
\addplot[ybar interval, fill=black!20!red, fill opacity=0.6, draw=black, area legend] table[row sep=crcr] {%
x	y\\
-2.5	0\\
-2.39795918367347	0\\
-2.29591836734694	0\\
-2.19387755102041	0\\
-2.09183673469388	0\\
-1.98979591836735	0\\
-1.88775510204082	0\\
-1.78571428571429	0\\
-1.68367346938776	0\\
-1.58163265306122	0\\
-1.47959183673469	0\\
-1.37755102040816	0\\
-1.27551020408163	0\\
-1.1734693877551	0\\
-1.07142857142857	0\\
-0.969387755102041	0\\
-0.86734693877551	0\\
-0.76530612244898	0\\
-0.663265306122449	0\\
-0.561224489795918	0\\
-0.459183673469388	0\\
-0.357142857142857	0\\
-0.255102040816327	0\\
-0.153061224489796	0\\
-0.0510204081632653	0\\
0.0510204081632653	0\\
0.153061224489796	0\\
0.255102040816327	0.228666666666667\\
0.357142857142857	2.51533333333334\\
0.459183673469388	3.98533333333334\\
0.561224489795918	1.176\\
0.663265306122449	0.424666666666667\\
0.76530612244898	0.555333333333334\\
0.86734693877551	0.424666666666667\\
0.969387755102041	0.228666666666666\\
1.07142857142857	0.0980000000000001\\
1.1734693877551	0\\
1.27551020408163	0.130666666666667\\
1.37755102040816	0.0326666666666666\\
1.47959183673469	0\\
1.58163265306122	0\\
1.68367346938776	0\\
1.78571428571429	0\\
1.88775510204082	0\\
1.98979591836735	0\\
2.09183673469388	0\\
2.19387755102041	0\\
2.29591836734694	0\\
2.39795918367347	0\\
2.5	0\\
};
\addlegendentry{Posterior samples}

\addplot [color=black, dashed, line width=2.0pt]
  table[row sep=crcr]{%
1	0\\
1	4\\
};
\addlegendentry{True value}

\end{axis}

\begin{axis}[%
width=0.606\figurewidth,
height=0.265\figureheight,
at={(0\figurewidth,0.368\figureheight)},
scale only axis,
xmin=-2.5,
xmax=2.5,
ymin=0,
ymax=4,
ytick={\empty},
ylabel style={font=\color{white!15!black}},
ylabel={PMMH},
axis background/.style={fill=white}
]
\addplot[ybar interval, fill=mycolor1, fill opacity=0.6, draw=black, area legend] table[row sep=crcr] {%
x	y\\
-2.5	0\\
-2.39795918367347	0\\
-2.29591836734694	0\\
-2.19387755102041	0\\
-2.09183673469388	0\\
-1.98979591836735	0\\
-1.88775510204082	0\\
-1.78571428571429	0\\
-1.68367346938776	0\\
-1.58163265306122	0\\
-1.47959183673469	0\\
-1.37755102040816	0\\
-1.27551020408163	0\\
-1.1734693877551	0\\
-1.07142857142857	0\\
-0.969387755102041	0\\
-0.86734693877551	0\\
-0.76530612244898	0\\
-0.663265306122449	0\\
-0.561224489795918	0\\
-0.459183673469388	0\\
-0.357142857142857	0\\
-0.255102040816327	0\\
-0.153061224489796	0\\
-0.0510204081632653	0\\
0.0510204081632653	0\\
0.153061224489796	0\\
0.255102040816327	0\\
0.357142857142857	0\\
0.459183673469388	0\\
0.561224489795918	0.000195998040019599\\
0.663265306122449	0.409537904620954\\
0.76530612244898	0.806433935660644\\
0.86734693877551	2.71369086309137\\
0.969387755102041	2.1310866891331\\
1.07142857142857	3.73905460945391\\
1.1734693877551	0\\
1.27551020408163	0\\
1.37755102040816	0\\
1.47959183673469	0\\
1.58163265306122	0\\
1.68367346938776	0\\
1.78571428571429	0\\
1.88775510204082	0\\
1.98979591836735	0\\
2.09183673469388	0\\
2.19387755102041	0\\
2.29591836734694	0\\
2.39795918367347	0\\
2.5	0\\
};
\addplot [color=black, dashed, line width=2.0pt, forget plot]
  table[row sep=crcr]{%
1	0\\
1	4\\
};
\end{axis}

\begin{axis}[%
width=0.606\figurewidth,
height=0.265\figureheight,
at={(0\figurewidth,0\figureheight)},
scale only axis,
xmin=-2.5,
xmax=2.5,
ymin=0,
ymax=10,
ytick={\empty},
ylabel style={font=\color{white!15!black}},
ylabel={SMC\textsuperscript{2}},
axis background/.style={fill=white}
]
\addplot[ybar interval, fill=black!50!green, fill opacity=0.6, draw=black, area legend] table[row sep=crcr] {%
x	y\\
-2.5	0\\
-2.39795918367347	0\\
-2.29591836734694	0\\
-2.19387755102041	0\\
-2.09183673469388	0\\
-1.98979591836735	0\\
-1.88775510204082	0\\
-1.78571428571429	0\\
-1.68367346938776	0\\
-1.58163265306122	0\\
-1.47959183673469	0\\
-1.37755102040816	0\\
-1.27551020408163	0\\
-1.1734693877551	0\\
-1.07142857142857	0\\
-0.969387755102041	0\\
-0.86734693877551	0\\
-0.76530612244898	0\\
-0.663265306122449	0\\
-0.561224489795918	0\\
-0.459183673469388	0\\
-0.357142857142857	0\\
-0.255102040816327	0\\
-0.153061224489796	0\\
-0.0510204081632653	0\\
0.0510204081632653	0\\
0.153061224489796	0\\
0.255102040816327	0\\
0.357142857142857	0\\
0.459183673469388	0\\
0.561224489795918	0\\
0.663265306122449	0\\
0.76530612244898	0\\
0.86734693877551	0\\
0.969387755102041	0\\
1.07142857142857	0\\
1.1734693877551	5.09600000000001\\
1.27551020408163	4.70400000000001\\
1.37755102040816	0\\
1.47959183673469	0\\
1.58163265306122	0\\
1.68367346938776	0\\
1.78571428571429	0\\
1.88775510204082	0\\
1.98979591836735	0\\
2.09183673469388	0\\
2.19387755102041	0\\
2.29591836734694	0\\
2.39795918367347	0\\
2.5	0\\
};
\addplot [color=black, dashed, line width=2.0pt, forget plot]
  table[row sep=crcr]{%
1	0\\
1	10\\
};
\end{axis}

\begin{axis}[%
width=0.262\figurewidth,
height=0.265\figureheight,
at={(0.689\figurewidth,0.735\figureheight)},
scale only axis,
xmin=0,
xmax=2.5,
ymin=0,
ymax=4,
ytick={\empty},
axis background/.style={fill=white},
title style={font=\bfseries},
title={$\theta_2$}
]
\addplot[ybar interval, fill=black!20!red, fill opacity=0.6, draw=black, area legend] table[row sep=crcr] {%
x	y\\
0	0\\
0.104166666666667	0\\
0.208333333333333	0\\
0.3125	0\\
0.416666666666667	0\\
0.520833333333333	0.096\\
0.625	0.096\\
0.729166666666667	0.224\\
0.833333333333333	0.224\\
0.9375	0.48\\
1.04166666666667	0.224\\
1.14583333333333	0.448\\
1.25	0.288\\
1.35416666666667	0.544000000000001\\
1.45833333333333	0.256\\
1.5625	0.735999999999999\\
1.66666666666667	0.896000000000001\\
1.77083333333333	2.304\\
1.875	0.735999999999999\\
1.97916666666667	0.608\\
2.08333333333333	0.768000000000001\\
2.1875	0.128\\
2.29166666666667	0.255999999999999\\
2.39583333333333	0.288\\
2.5	0.288\\
};
\addplot [color=black, dashed, line width=2.0pt, forget plot]
  table[row sep=crcr]{%
1	0\\
1	4\\
};
\end{axis}

\begin{axis}[%
width=0.262\figurewidth,
height=0.265\figureheight,
at={(0.689\figurewidth,0.368\figureheight)},
scale only axis,
xmin=0,
xmax=2.5,
ymin=0,
ymax=10,
ytick={\empty},
axis background/.style={fill=white}
]
\addplot[ybar interval, fill=mycolor1, fill opacity=0.6, draw=black, area legend] table[row sep=crcr] {%
x	y\\
0	0\\
0.104166666666667	0\\
0.208333333333333	0\\
0.3125	0\\
0.416666666666667	0\\
0.520833333333333	0\\
0.625	5.75015049849502\\
0.729166666666667	2.65792542074579\\
0.833333333333333	0.785944140558595\\
0.9375	0.00355196448035519\\
1.04166666666667	0.365468345316547\\
1.14583333333333	0.0342716572834271\\
1.25	0.000575994240057599\\
1.35416666666667	0\\
1.45833333333333	0.0010559894401056\\
1.5625	0.000383996160038399\\
1.66666666666667	0.0002879971200288\\
1.77083333333333	0.000383996160038399\\
1.875	0\\
1.97916666666667	0\\
2.08333333333333	0\\
2.1875	0\\
2.29166666666667	0\\
2.39583333333333	0\\
2.5	0\\
};
\addplot [color=black, dashed, line width=2.0pt, forget plot]
  table[row sep=crcr]{%
1	0\\
1	10\\
};
\end{axis}

\begin{axis}[%
width=0.262\figurewidth,
height=0.265\figureheight,
at={(0.689\figurewidth,0\figureheight)},
scale only axis,
xmin=0,
xmax=2.5,
ymin=0,
ymax=10,
ytick={\empty},
axis background/.style={fill=white}
]
\addplot[ybar interval, fill=black!50!green, fill opacity=0.6, draw=black, area legend] table[row sep=crcr] {%
x	y\\
0	0\\
0.104166666666667	0\\
0.208333333333333	0\\
0.3125	0\\
0.416666666666667	0\\
0.520833333333333	9.6\\
0.625	0\\
0.729166666666667	0\\
0.833333333333333	0\\
0.9375	0\\
1.04166666666667	0\\
1.14583333333333	0\\
1.25	0\\
1.35416666666667	0\\
1.45833333333333	0\\
1.5625	0\\
1.66666666666667	0\\
1.77083333333333	0\\
1.875	0\\
1.97916666666667	0\\
2.08333333333333	0\\
2.1875	0\\
2.29166666666667	0\\
2.39583333333333	0\\
2.5	0\\
};
\addplot [color=black, dashed, line width=2.0pt, forget plot]
  table[row sep=crcr]{%
1	0\\
1	10\\
};
\end{axis}
\end{tikzpicture}%

%% file: ex3.tex
%
%
\definecolor{mycolor1}{rgb}{0.00000,0.44700,0.74100}%
\begin{tikzpicture}

\begin{axis}[%
width=\figurewidth,
height=0.964\figureheight,
at={(0\figurewidth,0\figureheight)},
scale only axis,
xmin=10,
xmax=200,
xlabel style={font=\color{white!15!black}},
xlabel={$T$},
ymin=0,
ymax=200,
ylabel style={font=\color{white!15!black}},
ylabel={$P$},
axis background/.style={fill=white}
]
\addplot [color=mycolor1, line width=2.0pt, forget plot]
  table[row sep=crcr]{%
10	42\\
31	68\\
52	91\\
73	108\\
94	116\\
116	98\\
137	138\\
158	150\\
179	111\\
200	174\\
};
\addplot [color=black, dashed, line width=2.0pt, forget plot]
  table[row sep=crcr]{%
10	59.1246545651585\\
200	160.075345434842\\
};
\end{axis}
\end{tikzpicture}%